\documentclass[twocolumn,prd,superscriptaddress,preprintnumbers,amsmath,amssymb,nofootinbib]{revtex4}
\usepackage{amssymb}
\usepackage{slashed}
\usepackage{graphicx}
\usepackage{graphics}
\usepackage{epsfig}
\usepackage{color}
\usepackage{soul}
\usepackage{tabularx}
\usepackage{amsmath}
\usepackage{amsfonts}
\usepackage{amssymb}
\usepackage{hyperref}
\usepackage{microtype}

\newcommand{\be}{\begin{equation}}
\newcommand{\ee}{\end{equation}}
\newcommand{\bea}{\begin{eqnarray}}
\newcommand{\eea}{\end{eqnarray}}

\begin{document}
\author{Pietro Don\`a}
\email[]{pietro$\_$dona@fudan.edu.cn}
\affiliation{Department of Physics \& Center for Field Theory and Particle Physics, Fudan University, 200433 Shanghai, China}

\author{Astrid Eichhorn}
\email[]{a.eichhorn@imperial.ac.uk} 
\affiliation{Blackett Laboratory, Imperial College, London SW7 2AZ, United Kingdom}

\author{Peter Labus}
\email[]{plabus@sissa.it} 
\affiliation{International School for Advanced Studies, via Bonomea 265, 34136 Trieste, Italy\\
and INFN, Sezione di Trieste}

\author{Roberto Percacci}
\email[]{percacci@sissa.it} 
\affiliation{International School for Advanced Studies, via Bonomea 265, 34136 Trieste, Italy\\
and INFN, Sezione di Trieste}

\title{Asymptotic safety in an interacting system of gravity and scalar matter}

\begin{abstract}
Asymptotic safety is an attractive scenario for the dynamics of quantum spacetime. Here, we work from a phenomenologically motivated point of view and emphasize that a viable dynamics for quantum gravity in our universe must account for the existence of matter. In particular, we explore the scale-dependence of a scalar matter-gravity-vertex, and investigate whether an interacting fixed point exists for the so-defined Newton coupling. We find a viable fixed point in the pure-gravity system, disregarding scalar quantum fluctuations. We explore its extensions to the case with dynamical scalars, and find indications of asymptotic safety in the matter-gravity system.
We moreover examine the anomalous dimensions for different components of the metric fluctuations, and find significant differences between the transverse traceless and  scalar component.
\end{abstract}

\maketitle

\section{Introduction}
\subsection{Renormalization Group scale and bimetric structure of gravity}
The perturbative non-renormalizability of General Relativity means that, 
if we aim at a quantum field theoretic description of gravity, a nonperturbative route is necessary. An interacting fixed point of the Renormalization Group (RG) flow provides a notion of nonperturbative renormalizability, known as the asymptotic safety scenario \cite{Weinberg:1980gg}. Just as asymptotic freedom in non-Abelian gauge theories allows to define quantum field theories that are consistent and predictive at all scales, asymptotic safety could play the same role for quantum gravity, or, indeed, other gauge theories in $d=4$ dimensions \cite{Litim:2014uca, Litim:2015iea} or beyond \cite{Gies:2003ic}. 
Due to the interacting nature of the fixed point, one cannot quantize small metric fluctuations around a flat (or, e.g., cosmological) background. Instead quantum fluctuations of the metric can become arbitrarily large. This clearly suggests that the notion of a given background, available in the quantization of other gauge theories, is not present in gravity. This begs the question: What does it mean to construct a RG flow for quantum gravity? How to define coarse-graining, when spacetime itself, and therefore any measure of momentum-scales, is widely fluctuating?

The solution lies in the use of the background field method \cite{Abbott:1980hw}, where we split the full metric according to
\be
g_{\mu \nu} = \bar{g}_{\mu \kappa} {\rm exp}[ h^{.}_{\,\,.}]^{\kappa}_{\nu}.\label{exppara}
\ee
We then define the path-integral over all metric configurations as the path-integral over the fluctuation field $h_{\mu \nu}$, and the background metric $\bar{g}_{\mu \nu}$ can be used to set a scale. In particular, the fluctuation field can be decomposed into eigenfunctions of the background covariant Laplacian $-\bar{D}^2$, and those eigenmodes with large eigenvalues are declared to be the ``high-momentum" modes. In an RG flow from the ultraviolet (UV) to the infrared (IR), those modes are integrated out first.
It is important to realize that the fluctuation field can have arbitrary amplitude, so this split does not entail a perturbative treatment, and in fact varying $h_{\mu \nu}$ allows to reach every possible Riemannian metric $g_{\mu \nu}$ \cite{Demmel:2015zfa}.
The relation \eqref{exppara}, referred to as the exponential parameterization, has first been studied in the context of functional RG flows in a unimodular setting \cite{Eichhorn:2013xr,Eichhorn:2015bna}, and has also been argued to be superior to the linear parameterization in the context of standard gravity \cite{Nink:2014yya,Percacci:2015wwa,Demmel:2015zfa, Labus:2015ska,Ohta:2015efa,Gies:2015tca}.

In this framework, the effective action at a scale $k$, where degrees of freedom of momenta higher than $k$ (as determined by $\bar{g}_{\mu \nu}$) have been integrated out, depends on the background metric and the fluctuation metric, i.e., $\Gamma_k = \Gamma_k[\bar{g}_{\mu \nu}, h_{\mu \nu}]$. This dependence is such that one cannot recombine $\bar{g}_{\mu \nu}$ and $h_{\mu \nu}$ to give the full metric.
This ``split-symmetry" breaking is due to two sources:
The first is the gauge fixing term, which gauge-fixes the fluctuations with respect to the background, e.g., using a harmonic gauge condition $F_{\nu}=\bar{D}^{\mu}h_{\mu \nu} -\frac{1}{2} \bar{D}_{\nu}h^{\kappa}_{\kappa}$.
The second is the cutoff term, that is introduced into the path integral to implement a momentum-shell-wise integration: It acts as a mass-like term for fluctuations of low momenta and therefore has the structure $h_{\mu \nu} R^{\mu \nu \kappa \lambda} (y) h_{\kappa \lambda}$, where $y=-\bar{D}^2/k^2$ and $-\bar{D}^2$ denotes the background-covariant Laplacian\footnote{$R^{\mu \nu \kappa \lambda}$ should not be confused with the
Riemann tensor.}.
As a consequence, couplings of background operators and fluctuation operators do not share the same beta function. For instance, one can define a Newton coupling from the prefactor of the $\bar{R}$ term in the effective action, or from the momentum-squared part of the graviton three- point function or from a graviton-matter vertex. These three definitions of the Newton coupling obey a different Renormalization Group running. Modified Ward-identities govern the background-field dependence of the results, and have to be imposed on the RG flow, but work along these lines is still in its infancy.

In the literature on asymptotically safe gravity, many results are obtained within a single-metric approximation, where the difference between background couplings and fluctuation couplings is ignored. There, one finds an interacting fixed point with a finite number of relevant couplings, i.e., free parameters \cite{Reuter:1996cp,Dou:1997fg,Reuter:2001ag,Lauscher:2001ya,Lauscher:2002sq,Litim:2003vp,Fischer:2006fz,Machado:2007ea,Eichhorn:2009ah,
Codello:2006in,Codello:2008vh,Benedetti:2009rx,Eichhorn:2010tb, Groh:2010ta,Manrique:2011jc,Rechenberger:2012dt,Benedetti:2012dx,Dietz:2012ic,
Falls:2013bv,Benedetti:2013jk,Ohta:2013uca, Demmel:2014sga,Falls:2014tra, Falls:2015qga, Falls:2015cta,Gies:2015tca, Demmel:2015oqa}. For reviews, see \cite{ASreviews}.

First explorations of the bimetric structure in asymptotically safe quantum gravity have indicated that the evidence for asymptotic safety from the single-metric approximation is still present when resolving this approximation \cite{Manrique:2009uh,Manrique:2010mq,Manrique:2010am,Christiansen:2012rx, Codello:2013fpa,Christiansen:2014raa, Becker:2014qya,Christiansen:2015rva}.
One should note that at this stage, only few couplings have been considered in a bimetric setting, and higher-order truncations could yield different results.  As discussed in \cite{Bridle:2013sra} using the example of a scalar field, a single-metric approximation can result in spurious fixed points, and a treatment of the full bimetric structure is crucial. Within gravity-matter systems, a first step in this direction has been done in \cite{Dona:2013qba,Dona:2014pla}, where the anomalous dimension of the graviton and matter fields was evaluated in addition to the beta functions of the gravitational background couplings. In \cite{Christiansen:2014raa,Christiansen:2015rva}, RG flows formulated in terms of fluctuation field gravitational couplings have been investigated and lend quantitative support to the results for the pure-gravity case in the single-metric approximation. The system has been extended to include the effect of matter fluctuations on pure-gravity-couplings in \cite{Meibohm:2015twa}.
\subsection{Quantum spacetime and matter}
In this work, we will make a step forward in disentangling the running of fluctuation and of background couplings, focusing on the matter-gravity sector. From a phenomenological point of view, the matter-gravity couplings are particularly interesting for several reasons: First of all, these could become relevant in experimental tests of quantum gravity, e.g. in astrophysical or cosmological settings, where high enough energies to test quantum gravity effects might become reachable.
 Second, the existence of matter in our universe 
means that a quantum theory 
of spacetime by itself is not viable as a physical theory if it cannot accounted for matter as well.
While one could hope that matter ``emerges" as additional effective excitations of spacetime at low scales, it seems unlikely that the quantum dynamics of spacetime does indeed provide all observed matter degrees of freedom with the correct properties, as encoded in the intricate structure of the standard model. Thus, we will here follow the route towards a joint quantum theory of gravity and matter. In the context of asymptotic safety, this implies that a viable fixed point must exist not only for the gravitational interactions, but also for matter-gravity interactions and matter self-interactions. As the Standard Model by itself is most likely not asymptotically safe, the effect of gravity is conjectured to  induce a fixed point \cite{Shaposhnikov:2009pv}, see, e.g., for evidence in this direction \cite{Zanusso:2009bs,Vacca:2010mj, Harst:2011zx, Eichhorn:2011pc, Eichhorn:2012va,Oda:2015sma}. As a step towards showing that this could indeed be the case, we investigate the flow of a gravity-scalar-vertex, and show that it admits an interacting ultraviolet fixed point. We emphasize that the flow of this coupling is independent from the flow of the usual Newton coupling, defined with respect to gravitational vertices only. Our result therefore constitutes nontrivial evidence for the potential viability of asymptotic safety for a joint description of gravity and matter in our universe.

\section{Matter-gravity flows: Setup}

In the following we will analyze the Euclidean RG flow of the 1-graviton-2-scalar coupling. 
To derive its beta functions we will study the scale-dependence of the effective action $\Gamma_k$ which is governed by the Functional Renormalization Group equation \cite{Wetterich:1993yh, Morris:1993qb},  a.k.a. the Wetterich equation:
\be
\partial_t \Gamma_k = \frac{1}{2} {\rm STr} \left[\left( \Gamma_k^{(2)}+R_k\right)^{-1} \partial_t R_k \right].
\ee
The FRGE is formulated in terms of the dimensionless scale derivative $\partial_t = k\partial_k$, and $\Gamma_k^{(2)}$ denotes the second functional derivative of the flowing action with respect to the fields (a matrix in field space).
The supertrace $\rm STr$ includes a summation over the fields with an additional negative sign for Grassmannian fields, and a summation over the eigenvalues of the Laplacian in the kinetic term, that translates into a momentum-integral on a flat background. 
This equation depends on the full, field-dependent nonperturbative regularized propagator $\left(\Gamma_k^{(2)}+R_k \right)^{-1}$ which takes into account higher-loop effects while keeping a  rather simple  one loop form.  For reviews, see \cite{FRGreviews}.

\subsection{Truncation}
Our truncation consist in the Einstein-Hilbert action for the gravitational sector and a massless minimally coupled scalar field for the matter sector. To set it up, we start from an  auxiliary action $\hat{\Gamma}$ given in terms of the full metric $g_{\mu \nu}$, which reads
\be
\hat{\Gamma}= \Gamma_{\rm EH}+ S_{\rm gf} + \Gamma_{\rm kin}.
\ee
Herein
\be
\Gamma_{\rm EH} = -\frac{1}{16 \pi G} \int d^4x \sqrt{g} R\, 
\ee
and
\be
\Gamma_{\rm kin} = \frac{1}{2} \int d^4x \sqrt{g} g^{\mu \nu}\sum_{i=1}^{N_S}\partial_{\mu}\phi^i \partial_{\nu}\phi^i.
\ee
We drop a possible volume term in our calculation, as its fluctuations do not enter the RG flow in our choice of gauge, see below.

In particular, we will employ a York decomposition of the fluctuation field
\be
h_{\mu \nu} = h_{\mu \nu}^{TT} + \bar{D}_{\mu}v_{\nu}+ \bar{D}_{\nu}v_{\mu} + \bar{D}_{\mu}\bar{D}_{\nu}\sigma - \frac{1}{4}\bar{D}^2 \bar{g}_{\mu \nu}\sigma + \frac{1}{4}\bar{g}_{\mu \nu}h,
\ee
with $\bar{D}^{\mu}h_{\mu \nu}^{TT}=0=h^{\mu\, \, TT}_{\mu}$ and $\bar{D}^{\mu}v_{\mu}=0$
and  use the unimodular gauge, defined in \cite{Percacci:2015wwa}, which imposes a constant conformal mode, i.e.
\be
h=\rm const,
\ee
in the exponential parameterization. Moreover, vector fluctuations are also gauged to zero, leaving contributions of $h_{\mu \nu}^{TT}$and $\sigma$ to the running couplings. For details on the Faddeev-Popov ghost sector for this choice of gauge, see \cite{ercacci:2015wwa}.
We will employ a redefinition of the form $ \sigma \rightarrow \sqrt{(\bar{D}^{2})^2+\frac{4}{3}\bar{D}^{\mu}\bar{R}_{\mu \nu}\bar{D}^{\nu}} \sigma$, that cancels part of the Jacobian from the York decomposition \cite{Dou:1997fg}. 

To calculate the flow, we define a truncation in the following way: Starting from the action $\hat{\Gamma}$ we expand in powers of $h_{\mu \nu}$ up to fourth order, and then redefine $h_{\mu \nu} \rightarrow \sqrt{32 \pi G} h_{\mu\nu}$. In the nonperturbative FRG setting, the running of the prefactors of the terms at different order in $h_{\mu \nu}$ differs. We thus introduce several different ``avatars" of the Newton coupling, $G_3$, $G_4$, $g_3$, $g_4$ and $g_5$, and define our truncation to be
\bea
\label{gammarhs}
\Gamma_{k, \, rhs}&=&  \sqrt{\frac{G_3}{G}} \, \Gamma_{EH}^{(3,0)} + \frac{G_4}{G} \, \Gamma_{EH}^{(4,0)}\nonumber\\
&{}& + \sqrt{\frac{g_3}{G}} \, \Gamma_{\rm kin}^{(1,2)} + \frac{g_4}{ G} \, \Gamma_{\rm kin}^{(2,2)} + \left(\frac{g_5}{ G}\right)^{3/2} \, \Gamma_{\rm kin}^{(3,2)}\nonumber\\
&{}& + \mbox{ quadratic terms }.
\eea
Here $\Gamma^{(n,m)}$ stands for the terms of $n$-th order in the fluctuation field $h_{\mu \nu}$ 
and $m$-th order in the scalar field. 
Note that the action that contains the scalar fields is quadratic, so we only have terms with $m=0,2$.  Our redefinition of all separate prefactors of the different vertices allows us to explicitly distinguish these avatars of the Newton coupling, instead of approximating them all by $G$.  One should not expect a universal definition of the Newton coupling to exist in the nonperturbative quantum gravity regime, similarly to what has been found in the perturbative regime in \cite{Anber:2011ut}.
As replicas of Newton's coupling, $G_3$, $G_4$, $g_3$, $g_4$ and $g_5$ all have dimensionality $2-d$.
This justifies the different powers with which they appear
in the various terms in (\ref{gammarhs}).

In detail, the different terms on a flat background are given by: 
\bea
 \Gamma_{EH}^{(3,0)} &=&-\frac{1}{3!}
2\sqrt{32\pi G}
 \int d^4x\Bigl(\frac{3}{2} h_{\mu\nu}(\partial_{\mu}h_{\kappa\lambda})\partial_{\nu}h_{\kappa\lambda}\nonumber\\
&{}& - 3 h_{\kappa \lambda}(\partial_{\lambda}h_{\mu\nu})\partial_{\mu}h_{\nu\kappa}\Bigr),\label{GammaG3}\\
\Gamma_{EH}^{(4,0)}&=& - \frac{1}{4!}
64\pi G
\int d^4x\Bigl( 3h_{\mu\nu}h_{\kappa\lambda} (\partial_{\mu}h_{\kappa\rho})\partial_{\lambda}h_{\rho\nu}\nonumber\\
&+&h_{\nu\kappa}h_{\kappa\rho}(\partial_{\sigma}h_{\mu\nu})\partial_{\mu}h_{\rho\sigma}  
-3 h_{\mu\nu}h_{\nu\kappa}(\partial_{\kappa}h_{\rho\sigma})\partial_{\mu}h_{\rho\sigma}\nonumber\\
&+&4 h_{\mu\nu}h_{\nu\kappa}(\partial_{\mu}h_{\rho\sigma})\partial_{\sigma}h_{\rho\kappa}
-2 h_{\mu\nu}h_{\kappa \lambda}(\partial_{\lambda}h_{\nu\rho})\partial_{\rho}h_{\mu\kappa} \nonumber\\
&+& h_{\mu\nu}h_{\kappa\lambda}(\partial_{\rho}h_{\mu\kappa}) \partial_{\rho}h_{\nu\lambda} 
\nonumber\\
&-& h_{\mu\nu}h_{\nu\kappa}(\partial_{\rho}h_{\mu\sigma})\partial_{\rho}h_{\kappa\sigma}\Bigr),\label{GammaG4}\\
\Gamma_{kin}^{(1,2)} &=& - \frac{\sqrt{32 \pi  G}}{2} \int d^4x \sqrt{\bar{g}} h^{\mu \nu}\sum_{i=1}^{N_S}\partial_{\mu}\phi^i \partial_{\nu}\phi^i,\label{Gammag3}\\
\Gamma_{kin}^{(2,2)} &=& \frac{ 32 \pi  G}{2 } \int d^4x \sqrt{\bar{g}} h^{\mu \rho}h_{\rho}^{\phantom{\rho}\nu}\sum_{i=1}^{N_S}\partial_{\mu}\phi^i \partial_{\nu}\phi^i,\label{Gammag4}\\
\Gamma_{kin}^{(3,2)} &=& - \frac{(32 \pi  G)^{3/2}}{2}  \int d^4x \sqrt{\bar{g}} h^{\mu \rho}h_{\rho}^{\phantom{\rho}\lambda}h_{\lambda}^{\phantom{\lambda}\nu}\sum_{i=1}^{N_S}\partial_{\mu}\phi^i \partial_{\nu}\phi^i,\nonumber\\
&{}&\label{Gammag5}
\eea
where appropriate symmetrizations are understood implicitly, as $h_{\mu \nu} = h_{\nu \mu}$. Here, we have already imposed the gauge $h= \rm const$, and thus the trace of the fluctuation field can be dropped from the vertices. This simplifies the vertices considerably.

In an abuse of notation, we do not distinguish between dimensionful and dimensionless couplings, as all our beta functions will always be expressed in terms of the dimensionless couplings, only,  whereas all couplings in Eq.~\eqref{GammaG3}- \eqref{Gammag5} are still dimensionful.

By the subscript $rhs$ in \eqref{gammarhs} we indicate that this action is used to define the vertices and propagators that enter the Wetterich equation.  In other words, these are the fluctuation-field structures that \emph{induce} the Renormalization Group flow.
In this paper we will not calculate the running of all these couplings but only the beta function of $g_3$, and the wave-function renormalizations $Z_\Psi$, with $\Psi=({\rm TT},\sigma,S)$.
The wave-function renormalizations $Z_\Psi$ nevertheless couple into the beta-functions of the essential couplings in a nontrivial way via the anomalous dimensions
\bea
\eta_\Psi &=&-\partial_t\ln Z_\Psi\ .
\eea

The running of $G_3$ has been calculated recently in \cite{Meibohm:2015twa} using a linear parametrization of the metric
and a more conventional gauge.
We find that the simpler structure of the gravity-matter vertex
avoids some of the issues that are encountered with multi-graviton vertices, and in any case it is of interest to compare the results of different procedures.

The coupled nature of the Wetterich equation clearly prevents us from defining a closed truncation, in which we can extract the flow of all couplings that we have included on the right-hand side; thus approximations are necessary in which some couplings contribute to the running of others, but their running is not calculated. 
The remaining couplings in (\ref{gammarhs}) 
can accordingly be treated in various ways.
Since they enter in the flow equation for $g_3$,
it is better to avoid a truncation where they are set to zero.
Instead, they can be set equal to $g_3$, or treated as free parameters.
We will discuss different possible approximations with respect to these higher-order couplings below.
It is important to realize that if we were to restrict our truncation to $g_3$, and set all other couplings to zero, all but 
Fig.~\ref{threevertexdiagsg3} would vanish. On the other hand, the original action is diffeomorphism invariant, and accordingly a 1-graviton-2-scalar-vertex is necessarily accompanied by a 2-graviton-2-scalar vertex etc. It is therefore expected that an improved truncation takes the existence of these couplings - and the corresponding diagrams - into account. To close the system of couplings, we clearly have to choose an approximation, and we will mostly opt for the choice $g_4=g_3$, $g_5=g_3$ in the following (which is dictated by dimensionality). To check how useful this approximation is, we keep track of the couplings separately; however we will not evaluate the flows of the higher-order couplings.

Note an interesting difference of $\beta_{g_3}$ to the running of the background Newton coupling: As there is no closed scalar loop contributing to $\beta_{g_3}$, its only dependence on $N_S$ arises through the anomalous dimension $\eta_{\rm TT}$.

\subsection{Projection rules}
We use the transverse traceless mode $h_{\mu \nu}^{TT}$ (satisfying $h^{TT\; \mu}_{\;\;\;\;\mu}=0$, $\bar{D}^{\mu}h_{\mu \nu}^{TT}=0$) to define the matter-gravity coupling $g_3$. As the running of $g_3$ can unambigously be extracted on a flat background, we will focus on the choice $\bar{g}_{\mu \nu} = \delta_{\mu \nu}$ in the following.

To extract the running of $g_3$, we employ a projection rule as follows:
\be
\partial_t \sqrt{g_3} = \frac{8}{3} \frac{1}{\sqrt{32 \pi}}\left(p_{1\mu}p_{2\nu} \frac{\delta}{\delta h_{\mu \nu}^{TT}(p_3)} \frac{\delta}{\delta\phi(p_1)} \frac{\delta}{\delta\phi(p_2)} \partial_t \Gamma_k \right)\Big|_{(p^2)^2},\label{projectiong3}
\ee
where we use the symmetric configuration for the three momenta, such that an angle of $2\pi/3$ lies between them, and their absolute value is $|p_1| = |p_2|=|p_3| =p$.
Note that the functional derivative with respect to the TT mode generates the projector
\be
\label{projector}
\mathcal{P}^{\rm TT}_{\mu\nu\kappa\lambda}=
\frac{1}{2}\left(T_{\mu
\kappa}T_{\nu \lambda} 
+ T_{\mu\lambda}T_{\nu \kappa} \right) 
- \frac{1}{d-1}T_{\mu\nu}T_{\kappa\lambda},
\ee
where $T_{\mu \nu}=\delta_{\mu\nu}-p_{\mu}p_{\nu}/p^2$. 
As we are using the transverse traceless component of the graviton for the projection, there is no mixing with nonminimal couplings that arise from the diffeomorphism-invariant operator $\phi^2 R$, since the first variation of $R$ does not have a transverse traceless component  on a flat background. 
Furthermore, working with a transverse traceless external graviton mode also excludes an admixture of non-diffeomorphism invariant operators at the same order of momenta, which might be generated by the flow and which depend on the momenta of the graviton.

Any given vertex contains a large number of different tensor structures, and these are not necessarily all featuring the same running coupling. In particular, transverse traceless structures and scalar structures could be expected to exhibit prominent differences in their running. As a first step into this direction, we distinguish the wave-function renormalization for the TT mode and the $\sigma$ mode, $Z_{\rm TT}$ and $Z_{\sigma}$. On the other hand, we do not distinguish the couplings in the same fashion.

To extract the flow of the wave-function renormalizations, we define projection rules as follows:
\bea
\partial_t Z_S
&=& \left( \frac{\partial}{\partial p^2}\frac{\delta}{\delta \phi(-p)}\frac{\delta}{\delta \phi(p)} \partial_t \Gamma_k\right) ,\\ 
\partial_t Z_{\rm TT}
&=& \left( \frac{\partial}{\partial p^2} \frac{\mathcal P^{\rm TT}_{\mu \nu\kappa\lambda}(p)}{5} \frac{\delta^2}{\delta h^{TT}_{\mu \nu}(p) \delta h^{TT}_{\kappa \lambda}(-p)} \partial_t \Gamma_k\right),\\
\partial_t Z_{\sigma}
&=&-\frac{8}{3} \left( \frac{\partial}{\partial p^2}\frac{\delta}{\delta \sigma(-p)}\frac{\delta}{\delta \sigma(p)} \partial_t \Gamma_k\right),
\eea
where the right-hand sides are evaluated at $p=0$, and at vanishing external fields $h_{\mu \nu}^{TT}$, $\sigma$ and $\phi$.


\section{Results for $\eta_{\rm TT}$, $\eta_{\sigma}$ $\eta_S$ and $\beta_{g_3}$}
For our explicit results, we will employ a regulator shape function of the form $R_{\Psi k}\left(p^2\right) = Z_{\Psi k} (p^2-k^2) \theta(k^2-p^2)$, with the appropriate wave-function renormalization for all modes,  \cite{Litim:2001up}.

\subsection{Anomalous dimension for the graviton modes} 

\begin{figure}[!here]
\centering
\hspace{11mm} \includegraphics[width=0.33\linewidth,clip=true,trim=3cm 24cm 14cm 2cm]{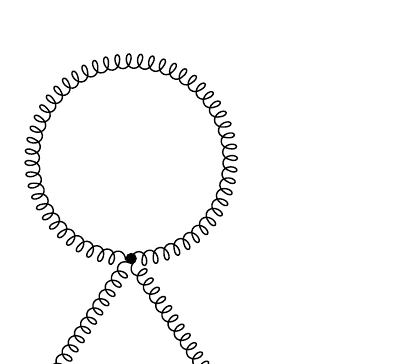} 
\hspace{9mm} \includegraphics[width=0.33\linewidth,clip=true,trim=3cm 24cm 14cm 2cm]{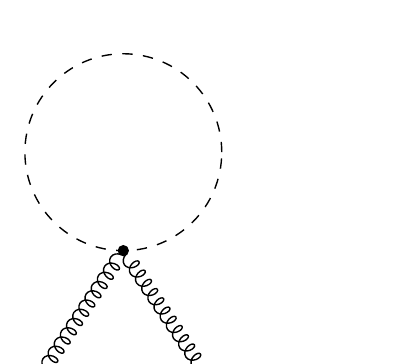}\newline
\includegraphics[width=0.45\linewidth,clip=true,trim=3cm 25cm 12cm 2cm]{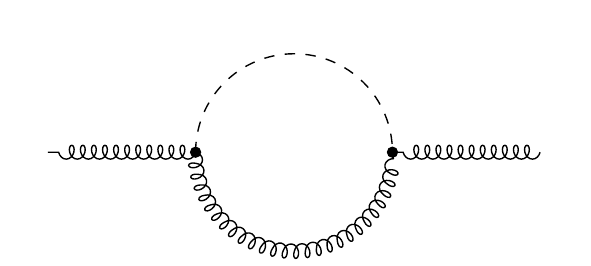} 
\includegraphics[width=0.45\linewidth,clip=true,trim=3cm 25cm 12cm 2cm]{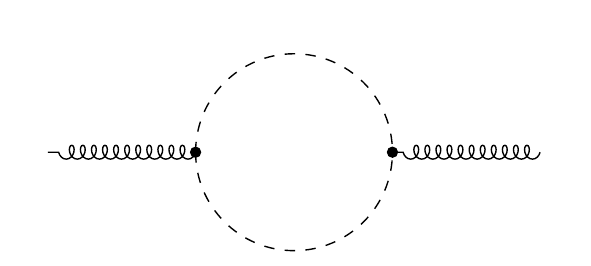}
\hspace{3mm} \includegraphics[width=0.45\linewidth,clip=true,trim=3cm 25cm 12cm 2cm]{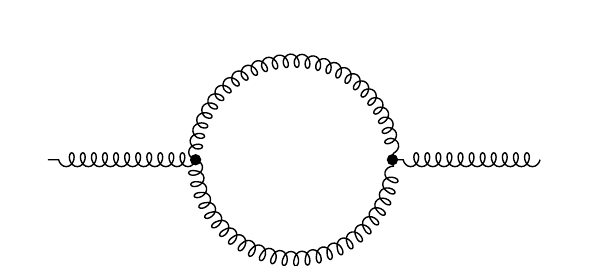}
\caption{\label{etahdiagsgrav} Metric diagrams contributing to the flow of $\eta_{\rm TT}$. Each diagram with n propagators occurs in n versions with the regulator insertion appearing on each of the n internal propagators in turn. Curly lines denote the transverse traceless metric mode and dashed lines the scalar graviton mode. Similar diagrams with external $\sigma$ lines contribute to the anomalous dimension $\eta_{\sigma}$.}
\end{figure}

\begin{figure}[!here]
\centering
\includegraphics[width=0.33\linewidth,clip=true,trim=3cm 24cm 14cm 2cm]{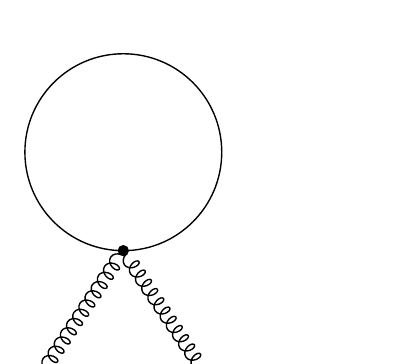}
\includegraphics[width=0.45\linewidth,clip=true,trim=3cm 25cm 12cm 2cm]{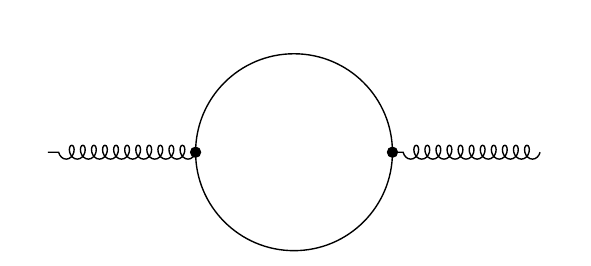}
\caption{\label{etahdiagsmatter} Matter diagrams contributing to the flow of $\eta_{\rm TT}$. Each diagram with n propagators occurs in n versions with the regulator insertion appearing on each of the n internal propagators in turn. Curly lines denote the transverse traceless metric mode and continuous lines the matter scalar. Similar diagrams with external $\sigma$ lines contribute to the anomalous dimension $\eta_{\sigma}$.}
\end{figure}

The purely metric diagrams contributing to $\eta_{\rm TT}$, cf.~Fig.~\ref{etahdiagsgrav} yield the following results.
\bea
\eta_{\rm TT} \big|_{\rm TT-tadpole} &=& \frac{145}{648 \pi}\, G_4\, (6-\eta_{\rm TT} ),\\
\eta_{\rm TT}  \big|_{\sigma- \rm tadpole} &=&  \frac{29}{324 \pi}\, G_4\, (-6+\eta_{\sigma}),\\
\eta_{\rm TT}  \big|_{TT,\,\sigma} &=&  \frac{25}{576 \pi}\, G_3 \, (16-\eta_{\rm TT}-\eta_{\sigma}),\label{TTsigmatwovertexetah}\\
\eta_{\rm TT}  \big|_{\sigma,\,\sigma} &=&  \frac{1}{216 \pi}\, G_3 \, (-31+5\eta_{\sigma}),\label{sigmatwovertexetah}\\
\eta_{\rm TT}  \big|_{TT,\,TT} &=&  \frac{5}{864 \pi}\, G_3 \, (-388+53\eta_{\rm TT} ).\label{TTtwovertexetah}
\eea
 The subscripts in (\ref{TTsigmatwovertexetah}) -- (\ref{TTtwovertexetah})
denote internal TT and $\sigma$ propagators in the two-vertex diagrams, respectively.
Similarly, there are two diagrams containing scalar fluctuations, one of them a tadpole (cf.~Fig.~\ref{etahdiagsmatter}) which vanishes due to the momentum-structure of the vertex.
\bea
\eta_{\rm TT}\big|_{\rm S-tadpole}  &=& 0,\\
\eta_{\rm TT}\big|_{\rm S,\,S}  &=& N_S \, \frac{1}{24 \pi} g_3.
\eea

Similarly, our results for the anomalous dimension of the $\sigma$ mode read
\bea
\eta_{\sigma} \big|_{\rm TT-tadpole} &=& \frac{55}{648 \pi}G_4(6-\eta_{\rm TT}),\\
\eta_{\sigma}  \big|_{\sigma- \rm tadpole} &=& \frac{11}{324 \pi}G_4 (-6+\eta_{\sigma}) ,\\
\eta_{\sigma}  \big|_{TT,\,\sigma} &=& \frac{5}{144\pi}G_3 (-16+\eta_{\rm TT}+ \eta_{\sigma}) ,\\
\eta_{\sigma}  \big|_{\sigma,\,\sigma} &=&\frac{1}{432\pi}G_3 (136 - 35 \eta_{\sigma}),\\
\eta_{\sigma}  \big|_{TT,\,TT} &=& \frac{5}{432\pi}G_3(40-23\eta_{\rm TT}) .
\eea

The matter contributions are given by
\bea
\eta_{\sigma}\big|_{\rm S-tadpole}  &=& 0,\\
\eta_{\sigma}\big|_{\rm S,\,S}  &=& N_S  \frac{1}{48 \pi}g_3 (8-3\eta_S).
\eea

In summary, we have 
\bea
\eta_{\rm TT}&=&N_S\frac{1}{24 \pi}g_3 + \frac{1}{1728 \pi} G_3 (-2928 + 455 \eta_{\rm TT} - 35 \eta_{\sigma}) \nonumber\\
&{}&- \frac{29}{648 \pi} G_4 (-18+5 \eta_{\rm TT} - 2 \eta_{\sigma}),\label{etaTTcomplete}\\
\eta_{\sigma}&=& N_S \frac{1}{48 \pi}(8-3 \eta_S)g_3 + \frac{1}{108 \pi} G_3 (24 - 25\eta_{\rm TT} - 5 \eta_{\sigma})\nonumber\\
&{}& +\frac{11}{648 \pi} G_4(18-5\eta_{\rm TT}+2\eta_{\sigma})
\label{etasigmacomplete}
\eea

Note that the sign of the matter contribution agrees for $\eta_{\rm TT}$ and $\eta_{\sigma}$ and is the opposite one from that in the linear parametrization and deDonder gauge, \cite{Dona:2013qba}. Since the exponential parametrization and the linear parametrization can be understood as underlying two distinct definitions of the configuration space for asymptotically safe quantum gravity, such a difference could possibly persist in extended truncations, and point towards a difference in the number of relevant directions in the two settings, see also \cite{Ohta:2015efa}.

The two-vertex diagrams enter with opposite signs in $\eta_{\rm TT}$ as compared to $\eta_{\sigma}$. This will imply that the two anomalous dimensions will typically have values of similar magnitude but opposite sign. Thus, setting $\eta_{\sigma} = \eta_{\rm TT}$ does not seem to be a good approximation, if indeed this trend persists beyond our truncation. Moreover, this could suggest that even in calculations without a York decomposition, it might be necessary to disentangle the tensor structures of the graviton, and work with projection tensors. Comparing to the anomalous dimension $\eta_h$ for the graviton (without York decomposition) in the linear parametrization, we observe that $\eta_{\rm TT}$ has the opposite, leading order \emph{negative} contribution from the pure-gravity fluctuations, cf.~Eq. 24 in \cite{Dona:2013qba}. 

\subsection{Anomalous dimension for the matter scalars}
\begin{figure}[t]
\centering
\hspace{11mm} \includegraphics[width=0.33\linewidth,clip=true,trim=3cm 24cm 14cm 2cm]{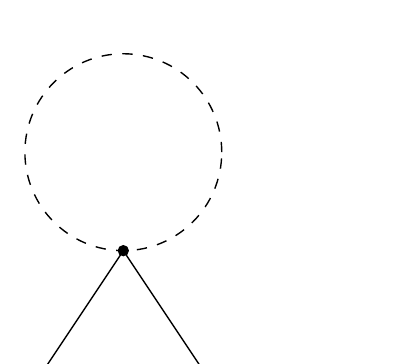}\quad
\hspace{9mm} \includegraphics[width=0.33\linewidth,clip=true,trim=3cm 24cm 14cm 2cm]{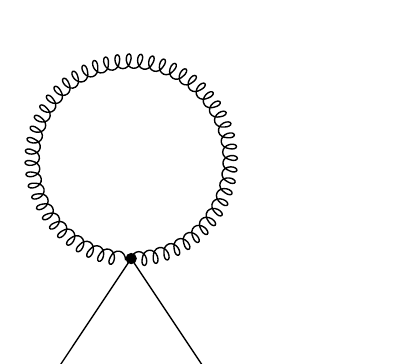}\newline
\includegraphics[width=0.45\linewidth,clip=true,trim=3cm 25cm 12cm 2cm]{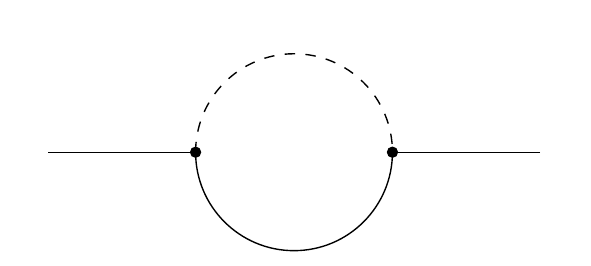}\quad
\includegraphics[width=0.45\linewidth,clip=true,trim=3cm 25cm 12cm 2cm]{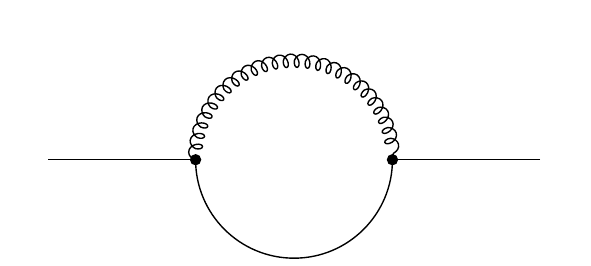}
\caption{\label{etasdiags}Metric diagrams contributing to the flow of $\eta_S$. Each diagram with n propagators occurs in n versions with the regulator insertion appearing on each of the n internal propagators in turn. Curly lines denote the transverse traceless metric mode, dashed lines the scalar graviton mode and continuous lines the scalar.}
\end{figure}

Similarly, tadpole diagrams and two-vertex diagrams contribute to the flow of $\eta_S$, cf.~Fig.~\ref{etasdiags}.
\bea
\eta_S\big|_{\rm TT-tadpole} &= &g_4\frac{5}{24\pi}\,(6-\eta_{\rm TT}), \label{etaS1}\\
\eta_S\big|_{\sigma- \rm tadpole} &= & g_4\frac{1}{12 \pi} \, (-6+\eta_{\sigma}),\label{etaS2}\\
\eta_S\big|_{\rm TT-two-vertex} &=& 0, \\
\eta_S\big|_{\sigma-\rm two-vertex} &=& g_3\frac{1}{16\pi} \, (16-\eta_{\sigma}-\eta_S)\label{etaS3}.
\eea

Overall, $\eta_S$ is positive at leading order, when $g_3>0, g_4>0$. This is the opposite behavior as that observed in the linear parameterization and the deDonder gauge, \cite{Dona:2013qba}.

\subsection{Flow of $\sqrt{g_3}$}

\begin{figure}[!here]
\includegraphics[width=0.45\linewidth,clip=true,trim=3cm 25cm 12cm 2cm]{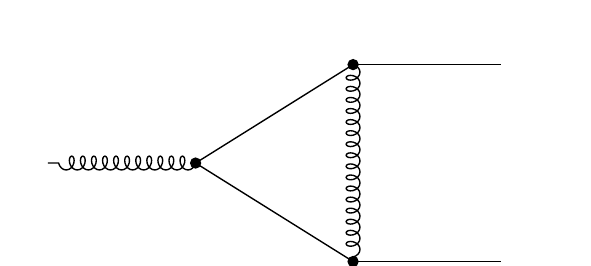}\quad \includegraphics[width=0.45\linewidth,clip=true,trim=3cm 25cm 12cm 2cm]{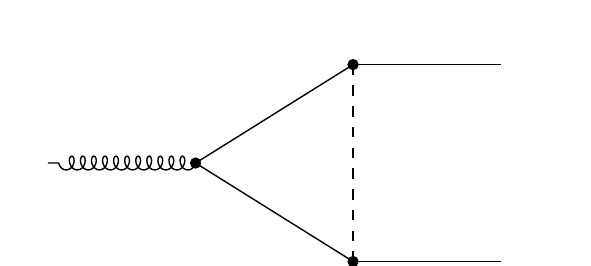}\newline
\caption{\label{threevertexdiagsg3} Three-vertex-diagrams contributing to the flow of $\sqrt{g_3}$, which feature only the coupling $\sqrt{g_3}$ itself at all vertices. Each diagram with n propagators occurs in n versions with the regulator insertion appearing on each of the n internal propagators in turn. Curly lines denote the transverse traceless metric mode, dashed lines the scalar graviton mode and continuous lines the scalar. }
\end{figure}

The flow of the two scalar-one graviton vertex
$\sqrt{g_3}$ is driven by three types of diagrams: First of all, there are three-vertex diagrams in which all vertices are $\sim \sqrt{g_3}$ themselves, cf.~Fig.~\ref{threevertexdiagsg3} (note that we do not distinguish between the coupling of two scalars and one $\sigma$ mode and the coupling of two scalars and the TT mode, although we use only the latter to read off the running of $g_3$.)

These three-vertex diagrams in Fig.~\ref{threevertexdiagsg3} only have a contribution from the $\sigma$ mode; the transverse traceless mode contributes to the running of the two-scalar-one-graviton-vertex at a higher order in the momenta, when our projection prescription
\eqref{projectiong3} is used. Note that the leading-order contribution to $\beta_{\sqrt{g_3}}$ comes with a positive sign. 

\bea
\beta_{\sqrt{g_3}}\big|_{\rm TT\, 3-vertex} &=& 0 \\
\beta_{\sqrt{g_3}}\big|_{\sigma \rm\, 3-vertex}  &=&   
\frac{1}{80 \pi} g_3^{3/2}\, (30-\eta_{\sigma}-2\eta_S)\label{betag3threevertexg3}
\eea

\begin{figure}[t]
\centering
\includegraphics[width=0.33\linewidth,clip=true,trim=3cm 23cm 14cm 2cm]{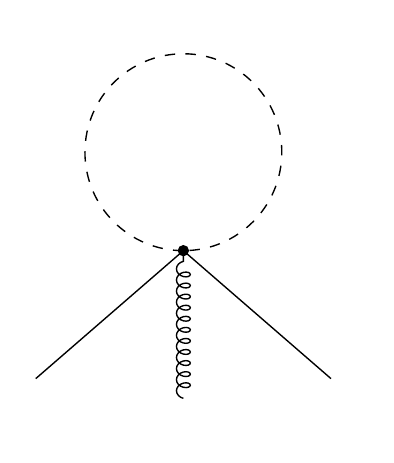}\quad \includegraphics[width=0.33\linewidth,clip=true,trim=3cm 23cm 14cm 2cm]{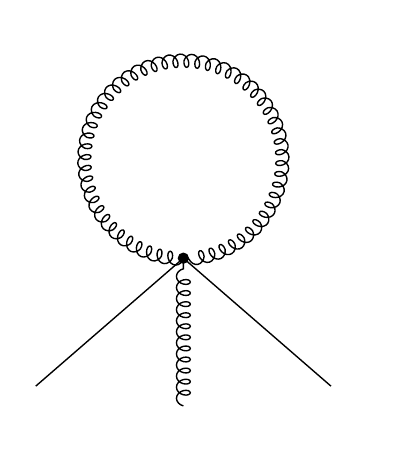}
\caption{\label{tadpolediags} Tadpole diagrams contributing to the flow of $\sqrt{g_3}$. Each diagram features a regulator insertion on each of the internal propagators, which we omit here. Curly lines denote the transverse traceless metric mode, dashed lines the scalar graviton mode and continuous lines the scalar.}
\end{figure}

\begin{figure}[t]
\includegraphics[width=0.45\linewidth,clip=true,trim=3cm 25cm 12cm 2cm]{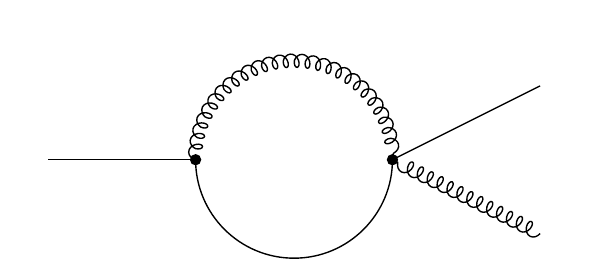}\quad 
\includegraphics[width=0.45\linewidth,clip=true,trim=3cm 25cm 12cm 2cm]{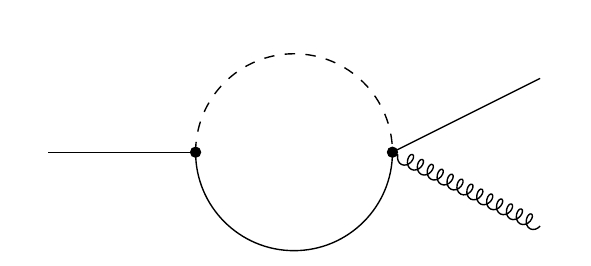}\newline
\includegraphics[width=0.45\linewidth,clip=true,trim=3cm 25cm 12cm 2cm]{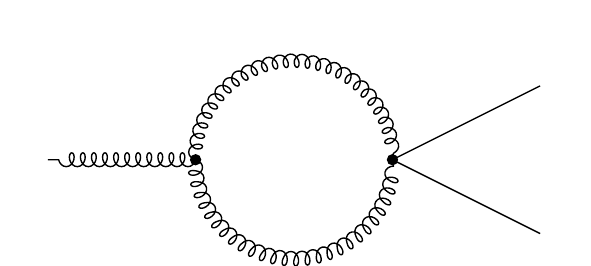}\quad 
\includegraphics[width=0.45\linewidth,clip=true,trim=3cm 25cm 12cm 2cm]{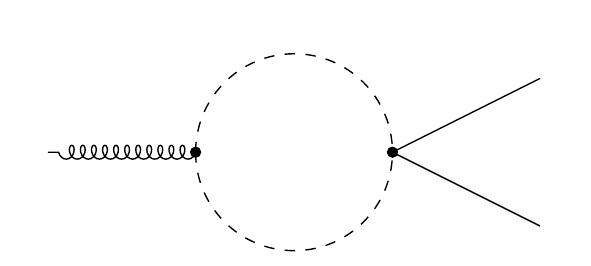}\quad
\includegraphics[width=0.45\linewidth,clip=true,trim=3cm 25cm 12cm 2cm]{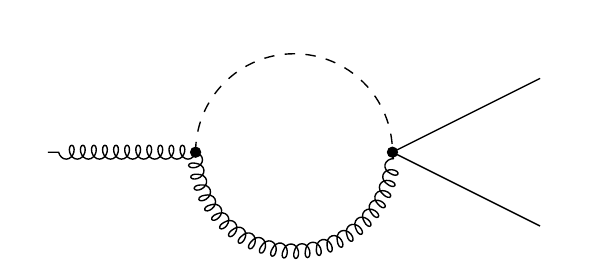}
\caption{\label{twovertexdiags} Two-vertex-diagrams contributing to the flow of $\sqrt{g_3}$. Each diagram occurs in two versions with the regulator insertion appearing on each of the two internal propagators in turn.Curly lines denote the transverse traceless metric mode, dashed lines the scalar graviton mode and continuous lines the scalar. The two diagrams in the first line feature only vertices that arise from the kinetic term for the scalar. The other three diagrams also feature pure-metric vertices which arise from the Einstein-Hilbert action.}
\end{figure}

Additionally, there are tadpole diagrams and two-vertex diagrams, which contain the couplings $g_4$ and $g_5^{3/2}$, which also arise from the kinetic term of the scalar, cf.~Fig.~\ref{tadpolediags} and Fig.~\ref{twovertexdiags}. 
The two tadpole diagrams contribute at $\mathcal{O}(g_5^{3/2})$ to the flow of $g_3$.
Equation \eqref{TTtadpole} denotes the transverse traceless graviton contribution, and 
\eqref{sigmatadpole} the scalar graviton contribution.
\bea
\beta_{\sqrt{g_3}}\Big|_{\rm TT\, tadpole} &=& 
-\frac{95}{648 \pi}g_5^{3/2} \, (6-\eta_{\rm TT}),\label{TTtadpole}\\
\beta_{\sqrt{g_3}}\Big|_{\sigma \rm\, tadpole} &=& 
\frac{19}{324 \pi} g_5^{3/2}\, (6-\eta_{\sigma}). \label{sigmatadpole}
\eea

The top two two-vertex diagrams in Fig.~\ref{twovertexdiags} contain only gravity-matter vertices, and contributes at $\mathcal{O}(\sqrt{g_3} g_4)$ to the flow of $\sqrt{g_3}$:
\bea
\beta_{\sqrt{g_3}}\Big|_{\substack{\rm TT,\,S\\\rm two-vertex}}  &=& 0,\\
\beta_{\sqrt{g_3}}\Big|_{\substack{\rm \sigma,\,S\\\rm two-vertex}} &= &-\frac{5}{72 \pi} \sqrt{g_3} g_4 \, (16-\eta_{\sigma}-\eta_S).
\eea

\begin{figure}[!top]
\centering
\includegraphics[width=0.45\linewidth,clip=true,trim=3cm 25cm 12cm 2cm]{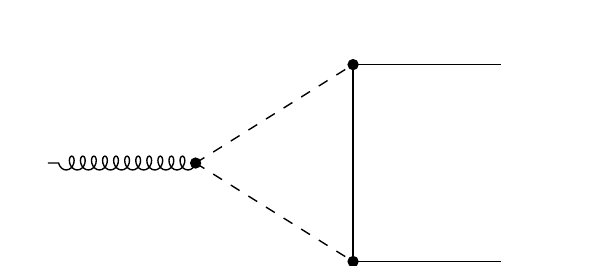} 
\includegraphics[width=0.45\linewidth,clip=true,trim=3cm 25cm 12cm 2cm]{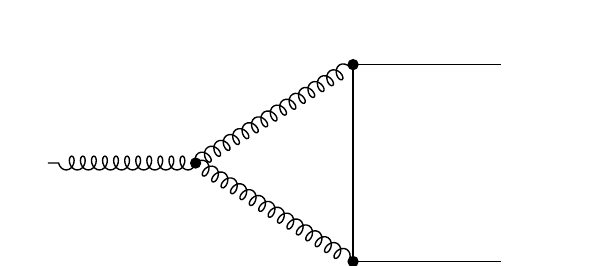}\newline
\includegraphics[width=0.45\linewidth,clip=true,trim=3cm 25cm 12cm 2cm]{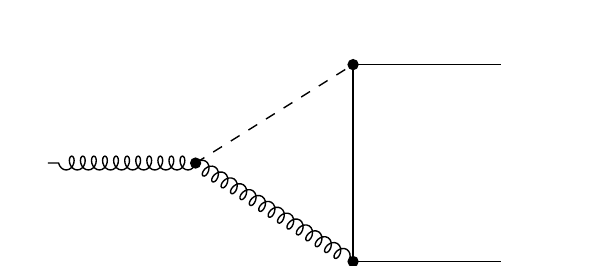}
\caption{\label{threevertexdiags} Three-vertex-diagrams contributing to the flow of $\sqrt{g_3}$, which also feature a pure-gravity vertex. Each diagram occurs in three versions with the regulator insertion appearing on each of the three internal propagators in turn. Curly lines denote the transverse traceless metric mode, dashed lines the scalar graviton mode and continuous lines the scalar. }
\end{figure}

Finally, there are two-vertex and three-vertex diagrams which also contain the vertex $\sqrt{G_3}$, which arises from the Einstein-Hilbert action, cf.~Fig.~\ref{twovertexdiags} and Fig.~\ref{threevertexdiags}. The lower three diagrams in Fig.~\ref{twovertexdiags} yield
\bea
\beta_{\sqrt{g_3}}\Big|_{\substack{\rm TT,\,TT\\\rm two-vertex}}   &=& -\frac{5}{216 \pi}  g_4 \sqrt{G_3}\, (8-\eta_{\rm TT}),\\
\beta_{\sqrt{g_3}}\Big|_{\substack{\rm \sigma,\,\sigma\\\rm two-vertex}}   &= &-\frac{1}{54 \pi} g_4 \sqrt{G_3} \, (-8 + \eta_{\sigma}), \\
\beta_{\sqrt{g_3}}\Big|_{\substack{\sigma \rm,\,TT\\\rm two-vertex}}  &=& 0.
\eea

The three-vertex diagrams with a pure-gravity vertex, cf.~Fig.~\ref{threevertexdiags}, yield the following contribution:

\bea
\beta_{\sqrt{g_3}}\Big|_{\substack{\sigma,\,\sigma,\,S \rm \\ \rm 3-vertex}} &=&\!\!\frac{1}{80\pi}g_3 \sqrt{G_3} \, (30-2\eta_{\sigma}-\eta_S),\\
\beta_{\sqrt{g_3}}\Big|_{\substack{\rm TT,\,TT,\,S\\ \rm 3-vertex}} &=& 0, \\
\beta_{\sqrt{g_3}}\Big|_{\substack{\sigma,\,\rm TT,\,S\\ \rm 3-vertex}} &=& 0.\label{threevertexpuregravityvertex}
\eea

\subsection{Beta function for $g_3$}
By summing Eq.~\eqref{betag3threevertexg3} to Eq.~\eqref{threevertexpuregravityvertex} we obtain the beta function for $\sqrt{g_3}$, from which we derive the beta function for the dimensionless $g_3$ as
\bea
\beta_{g_3}& =&
\left(2+\eta_{\rm TT}+2\eta_S\right)g_3
+\frac{3}{4\pi}g_3^2
+ \frac{3}{4\pi}g_3^{3/2}\sqrt{G_3} 
\nonumber\\
&-&\!\frac{20}{9\pi}g_3 g_4
- \frac{2}{3 \pi} \sqrt{g_3}\sqrt{G_3}g_4
- \frac{19}{18 \pi} g_5^{3/2}\sqrt{g_3}
\nonumber\\
&+& \!\left(\frac{5}{108\pi}g_4 \sqrt{G_3} 
+ \frac{95}{324\pi} g_5^{3/2}\right)\sqrt{g_3}\eta_{\rm TT}
\nonumber\\
&+&\! \!\Bigl(-\frac{1}{40 \pi}g_3^{3/2} 
- \frac{1}{20 \pi} g_3 \sqrt{G_3} 
+ \frac{5}{36 \pi} \sqrt{g_3} g_4 
\nonumber\\
&{}&+ \frac{1}{27 \pi}g_4 \sqrt{G_3} 
- \frac{19}{162\pi}g_5^{3/2}\Bigr)\sqrt{g_3}\eta_{\sigma}
\\
&+& \!\left(- \frac{1}{20 \pi}g_3^{3/2} 
- \frac{1}{40 \pi} g_3 \sqrt{G_3} 
+ \frac{5}{36 \pi}\sqrt{g_3} g_4 \right)\sqrt{g_3}\eta_S.
\nonumber
\eea
Herein, the factors $\eta_{\rm TT} g_3$ and $2 \eta_S g_3$ appear if the kinetic terms of both fields are redefined with a canonical prefactor, and the corresponding factors of the wave-function renormalization are absorbed in the coupling $g_3$.

 There is a significant difference to other possible definitions of a running Newton coupling: the beta function of $g_3$ depends implicitly on $N_S$, since only $\eta_{TT}$ and $\eta_{\sigma}$ contains a $N_S$ dependence.
There is no pure matter-loop contributing to $\beta_{g_3}$, though, as there is for some other definitions of a running Newton coupling, e.g., for the background Newton coupling \cite{Dona:2013qba}, or for the pure-gravity coupling $G_3$ \cite{Meibohm:2015twa}.

\section{Results for the pure gravity case}
Let us now analyze the fixed-point structure within various approximations.
If we consider the scalar as an external field, and only integrate out metric fluctuations, we can still consider $g_3$ as our definition of the running Newton coupling. In that case $\beta_{g_3}$ is determined by the tadpole diagrams in Fig.~\ref{tadpolediags} and the last three diagrams in Fig.~\ref{twovertexdiags}. 
The anomalous dimensions
only receive contributions from the diagrams in Fig.~\ref{etahdiagsgrav}
that do not contain matter in the loops.
They can be obtained from the general formulas simply putting $N_S=0$.
Here we further put the anomalous dimension $\eta_S$
to zero.

If we then employ the approximation $g_5 = g_4 = g_3$, $G_3 = G_4 = g_3$, we obtain 
\be
\beta_{g_3}= (2+\eta_{TT})g_3 
- \frac{31}{18 \pi}g_3^2 
+\frac{55}{162\pi}g_3^2\eta_{TT}
-\frac{13}{162\pi}g_3^2\eta_\sigma
\ee
where
\bea
\eta_{TT}&=&\frac{2g_3(1591 g_3-55296 \pi)}{2665 g_3^2-3384\pi g_3 +124416 \pi^2}
\\
\eta_\sigma&=&\frac{2g_3(16195 g_3+32832 \pi)}{2665 g_3^2-3384\pi g_3 +124416 \pi^2}
\eea

We  define the semi-perturbative approximation by setting $\eta_{\rm TT}= \eta_{\sigma}=\eta_S = 0$ on the right-hand sides of all diagrams contributing to the flow of the $\eta$'s. Then $\eta_{\rm TT}$ and $\eta_S$ still appear on the right-hand side of $\beta_{g_3}$. This semi-perturbative approximation removes potential poles from $\beta_{g_3}$ that are due to the nonperturbative structure of the $\eta$'s, and which could induce artificial zeros.
In this approximation
\bea
\beta_{g_3}= 2g_3 - \frac{47}{18 \pi}g_3^2 
-\frac{223}{648\pi}g_g^3.\label{purgravitysemipertbeta}
\eea
This structure, in particular the negative sign in front of the term $\sim g_3^2$, is similar to that found for other definitions of the Newton coupling.

The interplay between the dimensional term $2g_3$ and the leading order term from quantum fluctuations, $- g_3^2$, induces one real interacting fixed point as given in Tab.~\ref{puregravityFP_table}. The semi-perturbative approximation features another real fixed point, which we discard as a truncation artefact, as it is not present in the full beta function. Moreover, the perturbative approximation, in which we set all anomalous dimensions to zero everywhere, yields a similar result, where the critical exponent is of course set exactly by the negative dimensionality of the coupling. 

\begin{table}[!here]
\begin{tabular}{ccccc}
approximation&$g_{3\,\ast}$& $\theta$& $\eta_{\rm TT}$& $\eta_{\sigma}$\\ \hline \hline
full & 2.204 & 2.17 & -0.62 & 0.50\\ \hline
semi-pert.&2.203 & 2.17 & -0.62 & 0.37 \\ \hline
pert. ($\eta_{\rm TT}=0=\eta_{\sigma}$)& 3.65 & 2 & - & - \\\hline\hline
\end{tabular}
\caption{\label{puregravityFP_table}Coordinates and critical exponents at an interacting fixed point for vanishing scalar fluctuations.}
\end{table} 

The real part of the critical exponent is remarkably close to values in previous approximations, both in the single- and bi-metric case. We emphasize that this is a  rather non-trivial result. In our case, we define a coupling $g_3$, which is related to a gravity-scalar interaction vertex, in contrast to previous pure-gravity definitions. Accordingly, the diagrams entering the beta function have a fairly different structure, as does the beta function. It is rather reassuring to note that different ways of defining a Newton coupling and projecting the RG flow onto it result in similar universal properties.

The relatively large fixed-point value for $g_3$ is clearly responsible for the large absolute values of the anomalous dimensions, as $\eta \sim g_{3\,\ast}$: For instance, if we set $g_3=1$ by hand, we obtain $\eta_{\rm TT} = -0.28$ and $\eta_{\sigma} =0.20$. It has been observed previously that the exponential parameterization features a large fixed-point value for the Newton coupling in the single-metric approximation \cite{Percacci:2015wwa}, and our definition of the fluctuation-field coupling exhibits similar behavior.

The large negative value for the TT anomalous dimension suggests a propagator that decays with a higher power of the momentum in the UV, potentially suppressing the effect of TT quantum fluctuations in the UV.   On the other hand, the positive value for the scalar anomalous dimension implies that the $\sigma$ mode is actually enhanced in the UV. In particular, this could have very interesting consequences for gravity-operators at the UV fixed point. Operators of more ``scalar character" would be shifted towards relevance by the positive anomalous dimension $\eta_{\sigma}$, while operators with a larger contribution to the transverse traceless sector would be shifted towards irrelevance, even if the canonical dimension of both operators agrees. In particular, this could suggest that more complicated tensor structures, such as, e.g., powers of the Ricci tensor or Riemann tensor could be less relevant than their Ricci scalar counterparts.
(This concurs with an observation made in \cite{Codello:2006in}.)

Interestingly, the sign of $\eta_{\rm TT}$ is opposite to results in the linear parametrization \cite{Codello:2013fpa,Dona:2013qba,Christiansen:2014raa}.
If we identify $\eta_{\sigma} = \eta_{\rm TT}$ we obtain a fixed point with the properties listed in Tab.~\ref{puregravityFPetaapprox}.
We observe a comparable value for the critical exponent $\theta$ with  respect to the previous case. The anomalous dimension for the TT mode remains essentially the same. We see that the assumption $\eta_{\sigma}=\eta_{TT}$ even gives the wrong sign for $\eta_\sigma$. 
Since the anomalous dimensions contribute to the scaling dimensions of operators, this is of course a serious shortcoming.

\begin{table}[!here]
\begin{tabular}{cccc}
approximation&$g_{3\,\ast}$& $\theta$& $\eta_{\rm TT}$\\ \hline \hline
full & 2.20 & 2.19 & -0.67 \\\hline
semi-pert.&2.26 & 2.12 & -0.64 \\ \hline\hline
\end{tabular}
\caption{\label{puregravityFPetaapprox}Coordinates and critical exponents of an interacting fixed point with the approximation $\eta_{\sigma} = \eta_{\rm TT}$.}
\end{table} 

\begin{figure}[!here]
\includegraphics[width=\linewidth]{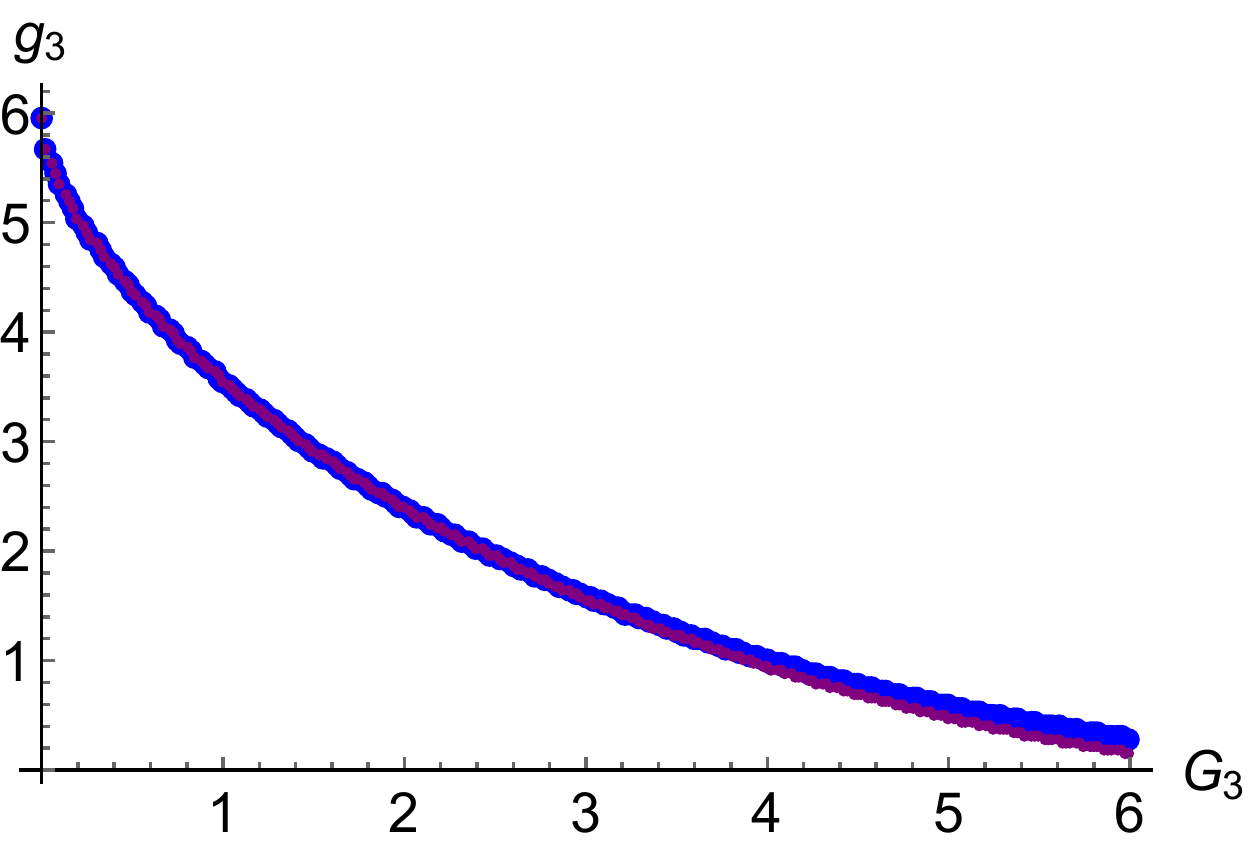}\\
\includegraphics[width=\linewidth]{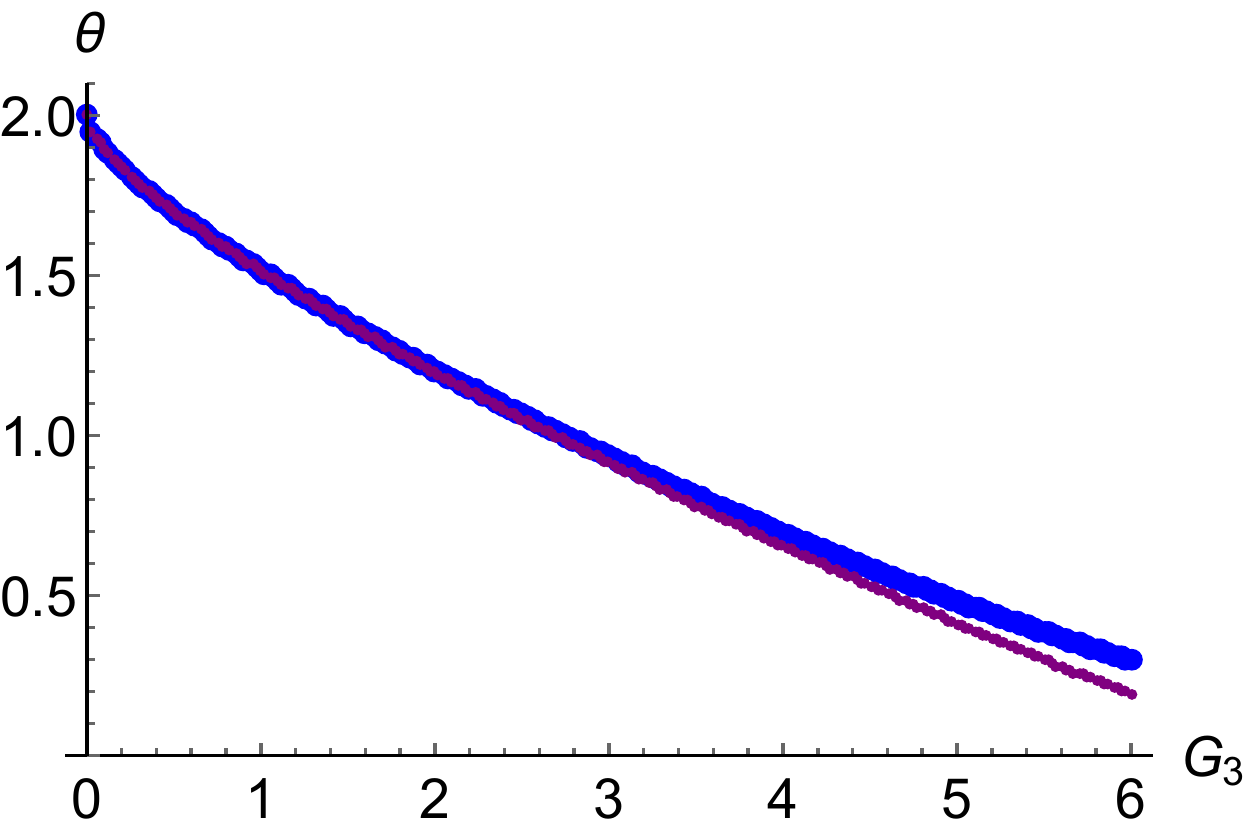}
\caption{\label{puregravityFP} We show the fixed point value for $g_3$ (upper panel) and the critical exponent $\theta$ (lower panel) as a function of $G_3=G_4$. The larger blue dots denote the full result and the smaller purple dots the semi-perturbative approximation.}
\end{figure}

To test the stability of our results with respect to extended truncations which would include separate beta functions for $G_3$ etc., we consider an approximation where $g_5=g_4=g_3$. We treat the pure-gravity coupling $G_4=G_3$ as an external parameter, and test whether a viable fixed point exists for values of these couplings between 0 and 6. 
For this case, we display the semi-perturbative result, as it allows a clear understanding of the terms:
\bea
\beta_{g_3} &=&2 g_3 - \frac{8}{9 \pi} g_3 G_3 
- \frac{19}{18 \pi}g_3^2
\\
&{}& - \frac{209 g_3^2 G_3}{648 \pi^2} 
- \frac{2}{3 \pi}g_3^{3/2}\sqrt{G_3}
- \frac{7}{324 \pi^2}g_3^{3/2}G_3^{3/2}\ .
\nonumber
\eea
When $G_3$ increases, the fixed point for $g_3$ decreases,
cf.~Fig.~\ref{puregravityFP}. 
We see that the fixed point that we have observed in the approximation $G_3=g_3$, persists for a large range of values of $G_3$.
We interpret this as a sign of stability.

\begin{figure}[!here]
\includegraphics[width=0.45\linewidth]{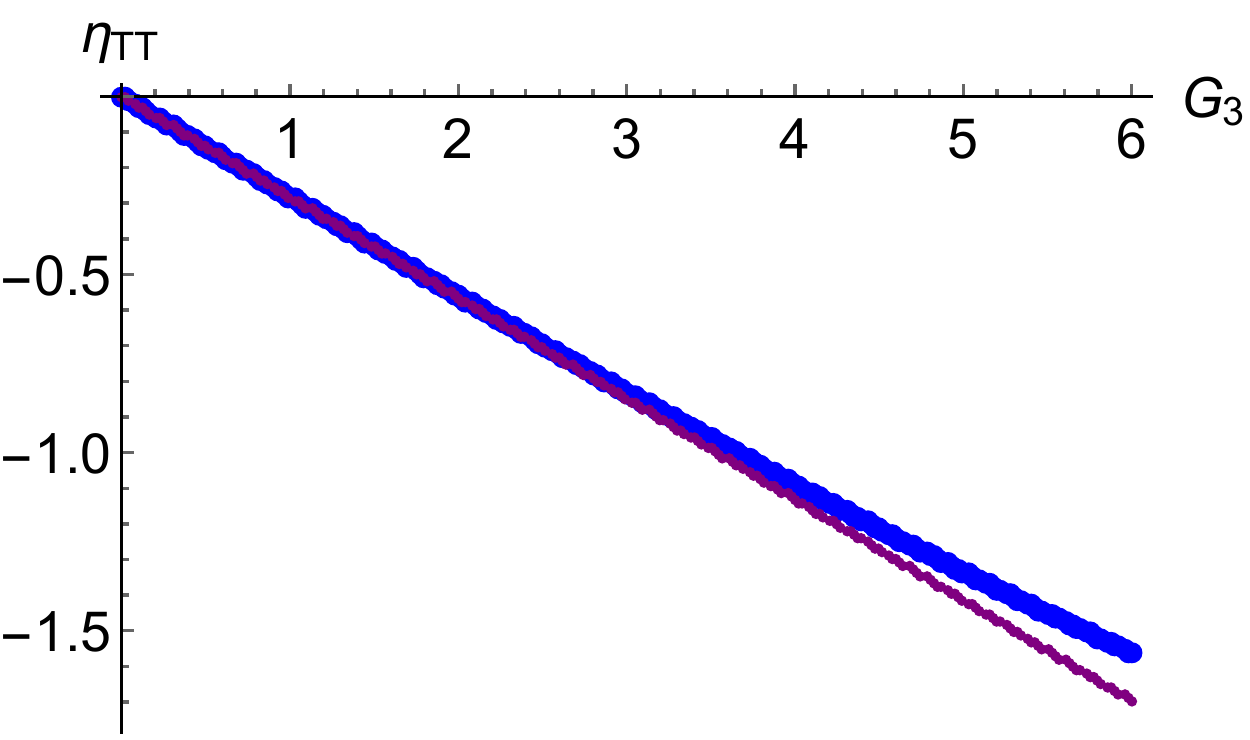}\quad
\includegraphics[width=0.45\linewidth]{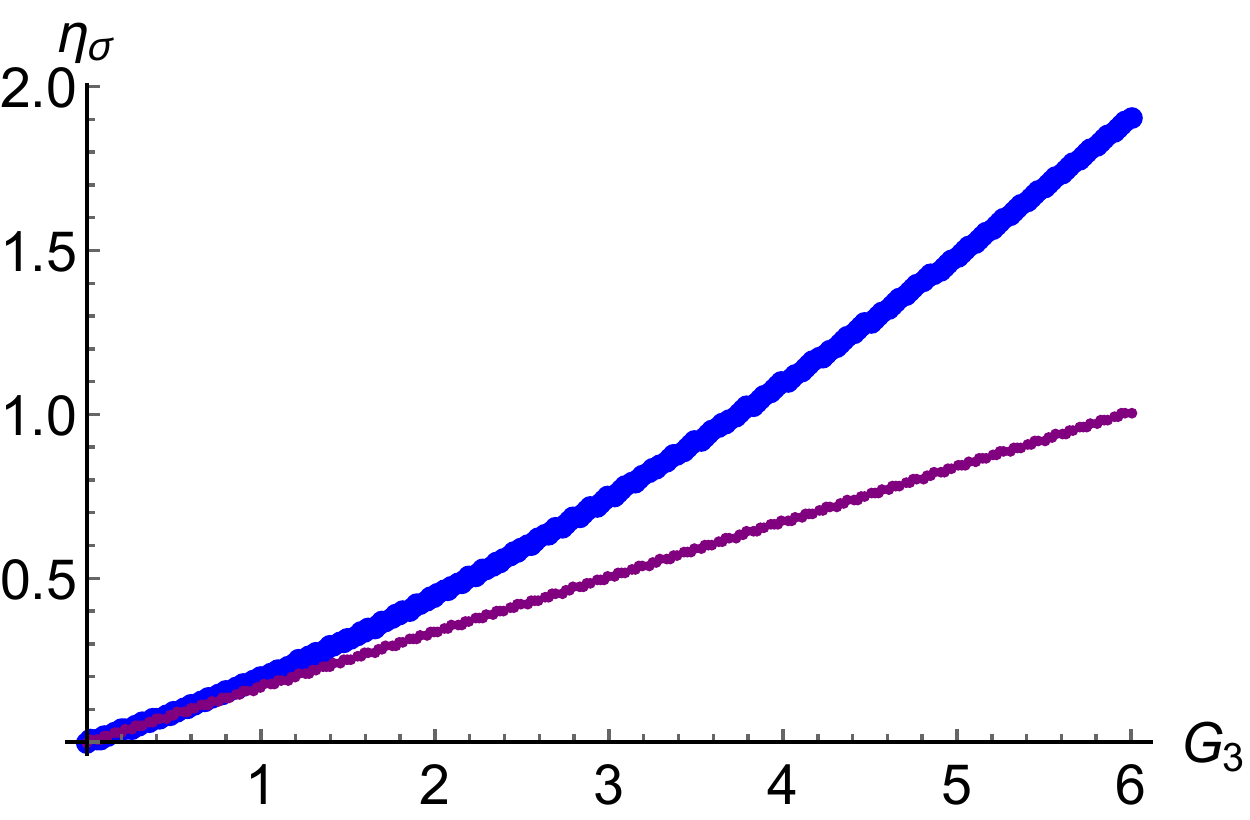}
\caption{\label{puregravityeta}We show the fixed point value for $\eta_{\rm TT}$ (left panel) and $\eta_{\sigma}$ (right panel) as a function of $G_3=G_4$. The larger blue dots denote the full result and the smaller purple dots the perturbative approximation.}
\end{figure}

Fig.~\ref{puregravityeta} shows the anomalous dimensions at the fixed point as functions of $G_3$. 
Note that they vanish for $G_3=0$, since,
in the absence of matter self interactions, they are 
generated only by diagrams proportional to $G_3$.
 Note that to obtain this result, the identification $g_5=g_3$, suggested by diffeomorphism invariance, is crucial, as it gives rise to the above structure of the beta function.

The critical exponent $\theta$ in this approximation is not equal to the result of Tab.~\ref{puregravityFP_table} at $G_3=g_3$, since $\partial \beta_{g_3}/\partial G_3 \vert_{G_3=g_3}$ contributes to the critical exponent quoted in that table. When we distinguish $g_3$ and $G_3$, then  $\partial \beta_{g_3}/\partial G_3 \vert_{G_3=g_3}$ yields an \emph{off-diagonal} contribution to the stability matrix, which can contribute to the critical exponents if operators mix at the non-Gau\ss{}ian fixed point. As we only evaluate the diagonal entry of the stabilty matrix in our approximation, where $G_3$ is treated as an external parameter, that contribution is absent.

We now supplement our beta functions by a beta function for the background Newton coupling, as obtained in \cite{Percacci:2015wwa}, where
\be
\beta_{\bar{G}}=2 \bar{G} - \frac{\bar{G}^2}{\pi} \left(\frac{15}{8}- \frac{5 \eta_{\rm TT}}{18 }+ \frac{\eta_{\sigma}}{24} \right).
\ee
We observe that $\eta_{\rm TT}$ and $\eta_{\sigma}$ enter with the opposite sign. At the fluctuation-field fixed point, $\eta_{\rm TT}<0$  and $\eta_{\sigma}>0$. This sign combination strengthens the gravitational fluctuation effects that induce a fixed point, lending further support to the observation that all modes of the graviton act towards asymptotic safety  \cite{Reuter:2008qx, Reuter:2008wj}.

Plugging in the fluctuation field fixed point-values - which are of course independent of $\bar{G}$, as they should - we obtain $\bar{G}_{\ast} = 3.04$ and $\theta=2$.

\section{Results for the interacting matter-gravity system}
In the following, we switch on scalar fluctuations, which adds several classes of diagrams to $\beta_{g_3}$ and additional contributions to $\beta_{g_3}$, $\eta_{\rm TT}$ and $\eta_{\sigma}$. Our main goal is to find out whether the pure-gravity results discussed above can be extended in a stable way to $N_S>0$, or whether scalars have a significant effect on the fixed point in our approximation. First, we will set $\eta_S=0$ by hand, while taking into account $\eta_{\rm TT}$ and $\eta_{\sigma}$. The reason for this unequal treatment of fluctuation fields will become clear below.

\subsection{Fixed-point results without scalar anomalous dimension}
As a first approximation, we set $G_3=G_4=g_4=g_5=g_3$, and find an extension of the pure-gravity fixed point to $N_S>0$. Unlike the beta functions for the background Newton coupling and the pure-gravity couplings $G_3$ and $G_4$, $\beta_{g_3}$ receives no correction from diagrams containing a scalar loop  in our approximation. Consequently, $\beta_{g_3}$ does not depend on $N_S$ explicitly, if we set all anomalous dimensions to zero. 
It only depends on $N_S$ if we include the anomalous dimensions $\eta_{\rm TT}$ and $\eta_{\sigma}$.
This gives rise to a beta function of the form
\be
\beta_{g_3}= 2g_3 -\frac{1}{3\pi}g_3^2\left(10-\frac{N_S}{8}\right)
+\mathcal{O}(g_3^3).
\label{norma}
\ee
Herein, the factor $10/3\pi$ differs from the  factor $47/18\pi$ in Eq.~\eqref{purgravitysemipertbeta} since it includes contributions form scalar fluctuations which do not scale with $N_S$.
In our determination of fixed points we also take into account all higher-order terms, but the main effect of scalars is clear from the $\mathcal{O}(g_3^2)$ term: As the contribution of scalars, that scales with $N_S$ explicitly, comes with the opposite sign from the asymptotic safety-inducing term which enters with $-10g_3^2/3\pi$, scalars push the fixed-point of $g_3$ towards larger values. This is the same effect that we have already observed for the background-system \cite{Dona:2013qba}. Interestingly, the contribution of scalars to the running of the three-graviton coupling features an opposite sign in the approximation used in \cite{Meibohm:2015twa}. On the other hand, the complete fixed-point dynamics in \cite{Meibohm:2015twa} is similar, as $G_{3\,\ast}$ is also pushed to larger values, since destabilizing effects of scalars show up in the momentum-independent part of the graviton-three- and two-point function.

\begin{figure}[!here]
\includegraphics[width=\linewidth]{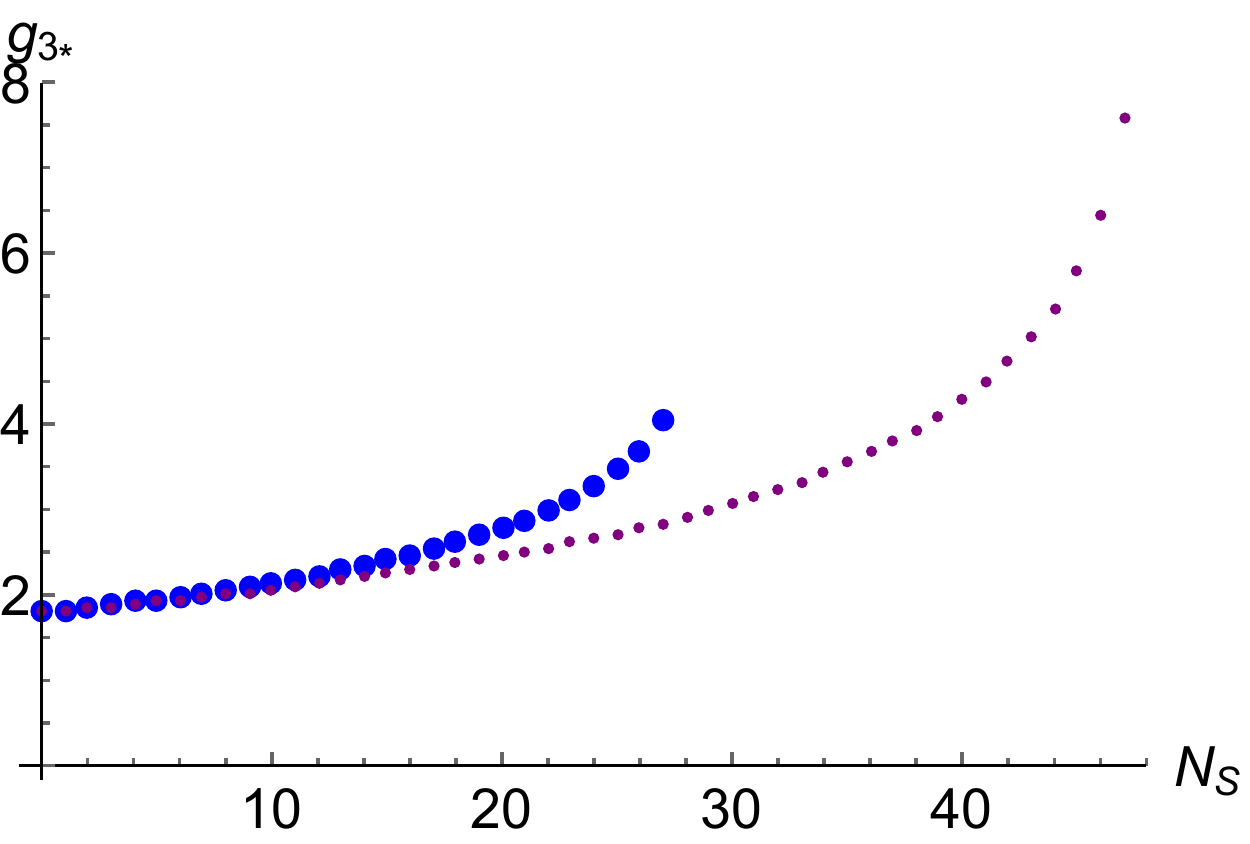}\\
\includegraphics[width=\linewidth]{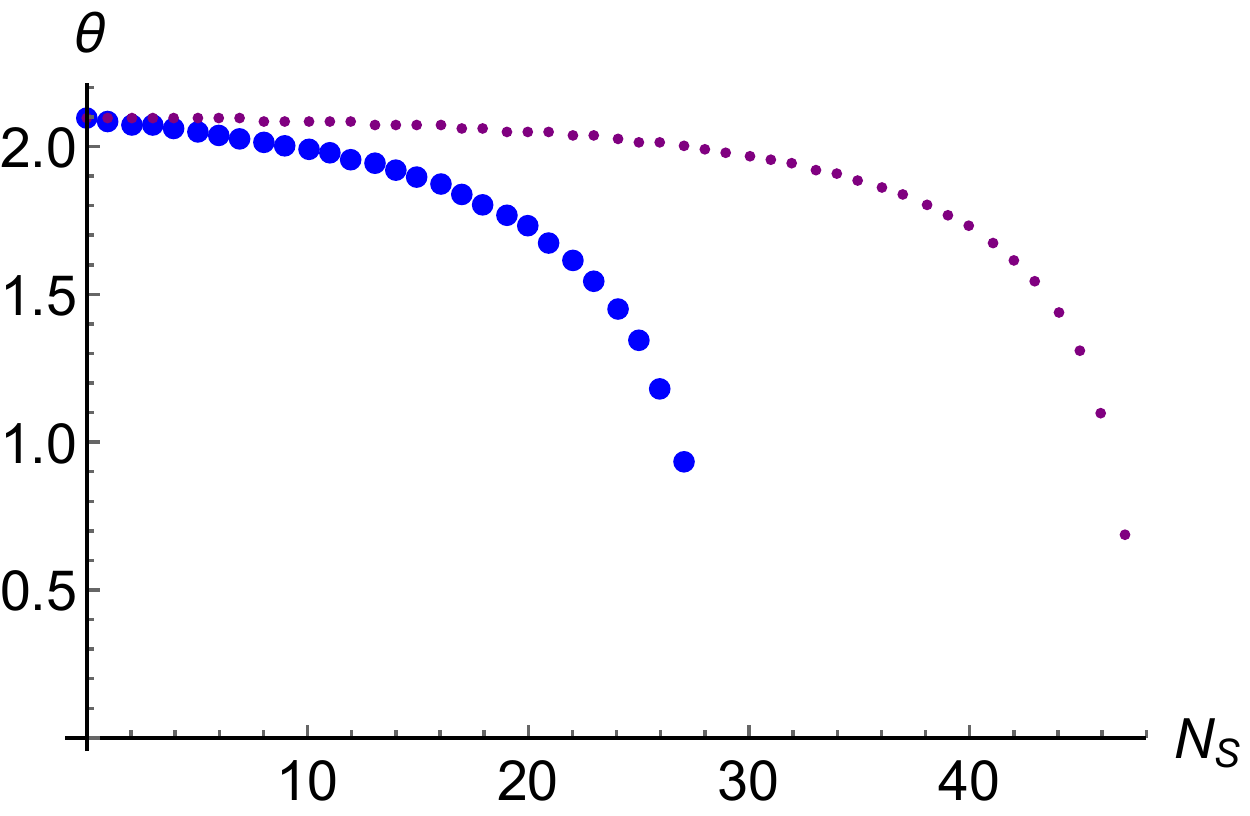}
\caption{\label{g3Nsnoetas}We show the fixed point value for $g_3$ (upper panel) and the critical exponent $\theta$ (lower panel) as a function of $N_S$. The larger blue dots denote the full result and the smaller purple dots the semi-perturbative approximation.}
\end{figure}

In our approximation, the fixed point merges with another, artificial fixed point at $N_S\approx 28$, where the fixed point disappear into the complex plane, cf.~Fig.~\ref{g3Nsnoetas}. While this could indicate the existence of a bound on $N_S$ in asymptotically safe gravity, we note that $\eta_{\sigma}>2$ already at $N_S=14$, indicating that a larger truncation is required to investigate that regime in more detail, see \cite{Meibohm:2015twa}. In the semi-perturbative approximation, the anomalous dimensions are slightly smaller, and the fixed-point collision accordingly occurs at larger $N_S$. 

\begin{figure}[!here]
\includegraphics[width=0.45\linewidth]{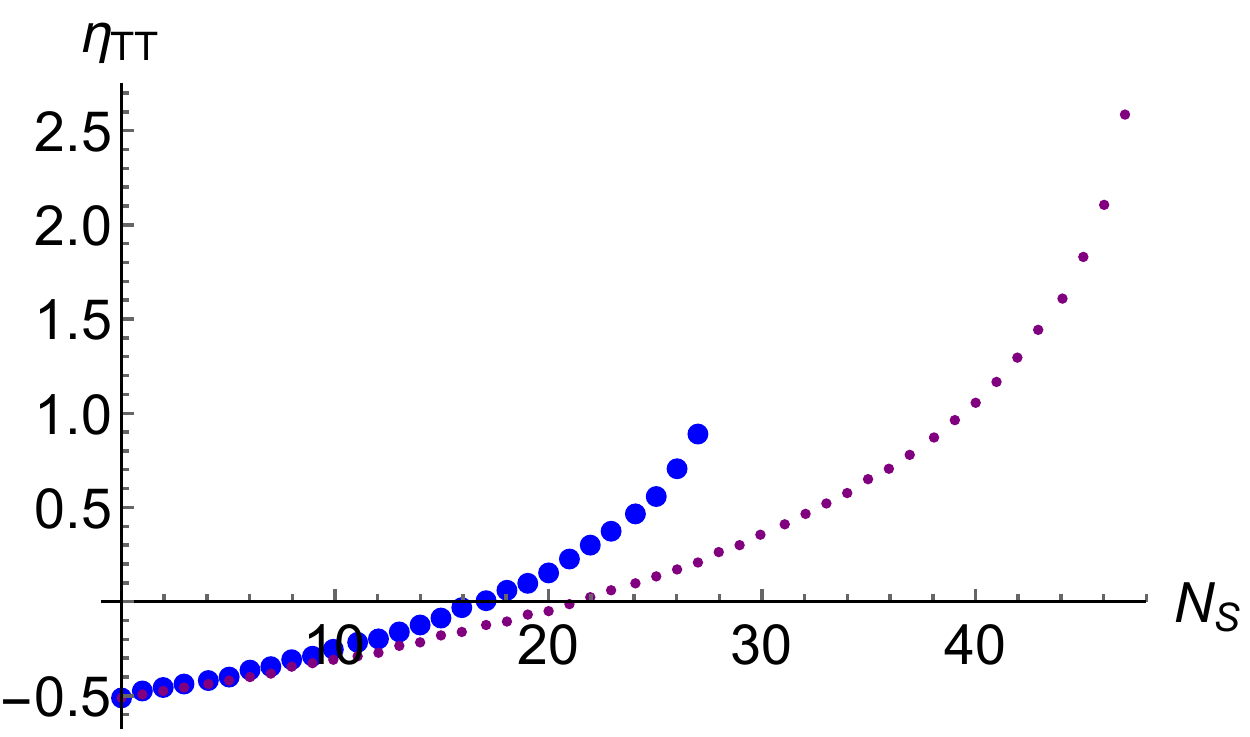}\quad
\includegraphics[width=0.45\linewidth]{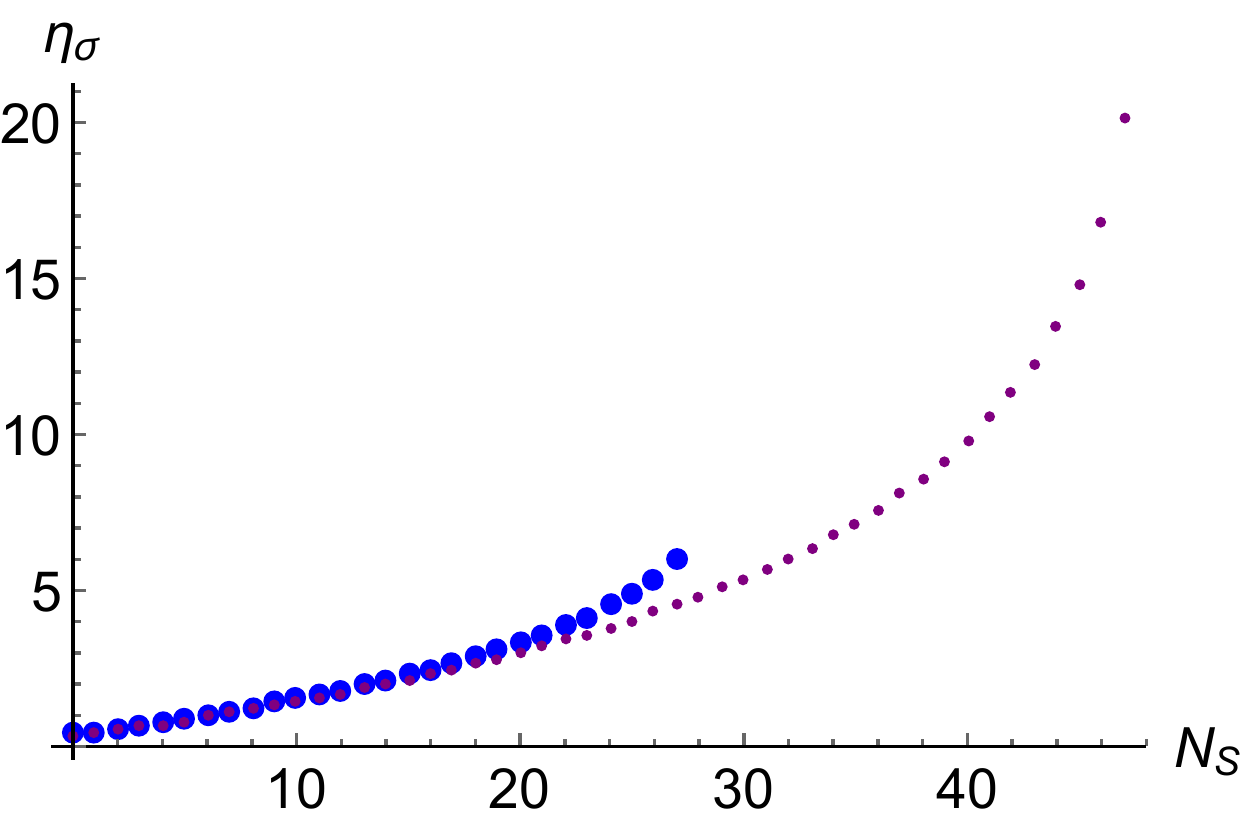}
\caption{\label{etaNsnoetas}We show the fixed point value for $\eta_{\rm TT}$ (left panel) and $\eta_{\sigma}$ (right panel) as a function of $N_S$. The larger blue dots denote the full result and the smaller purple dots the semi-perturbative approximation.}
\end{figure}

\begin{figure}[!here]
\includegraphics[width=\linewidth]{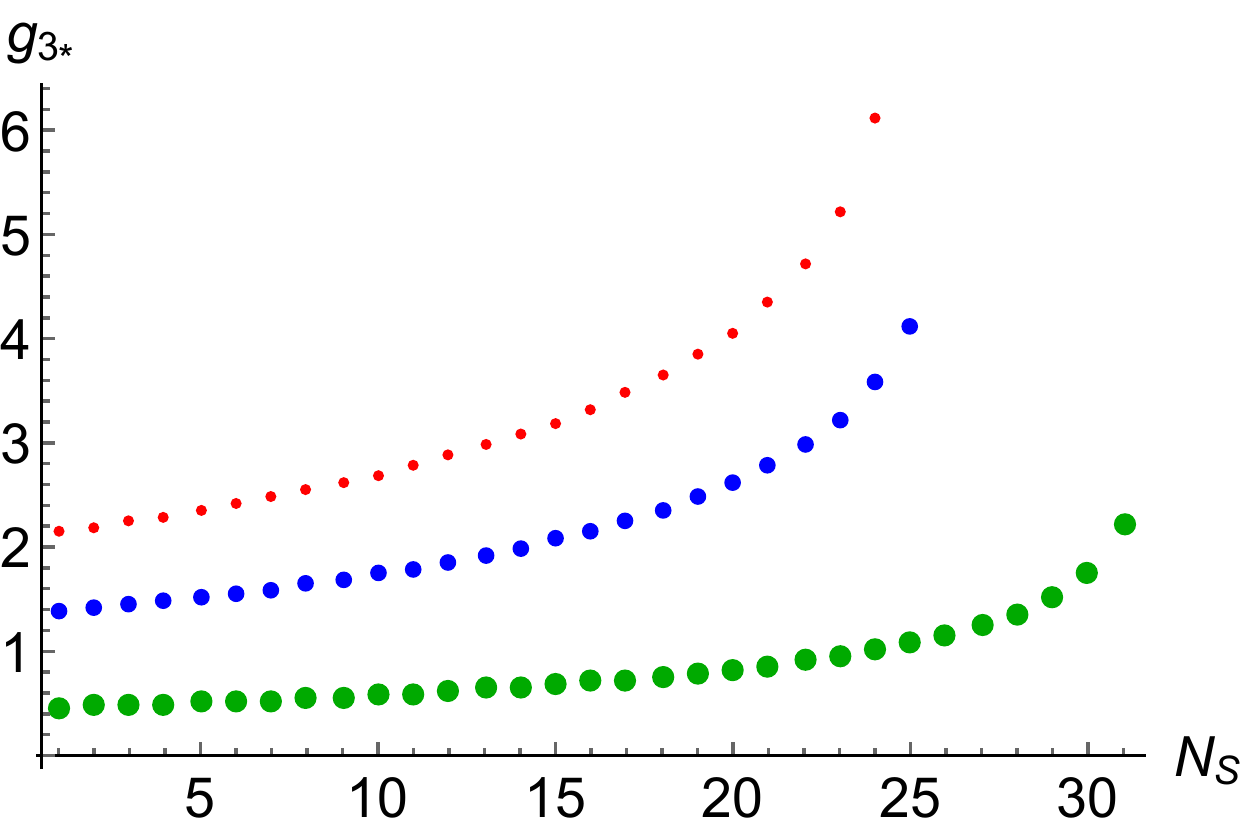}\\
\includegraphics[width=\linewidth]{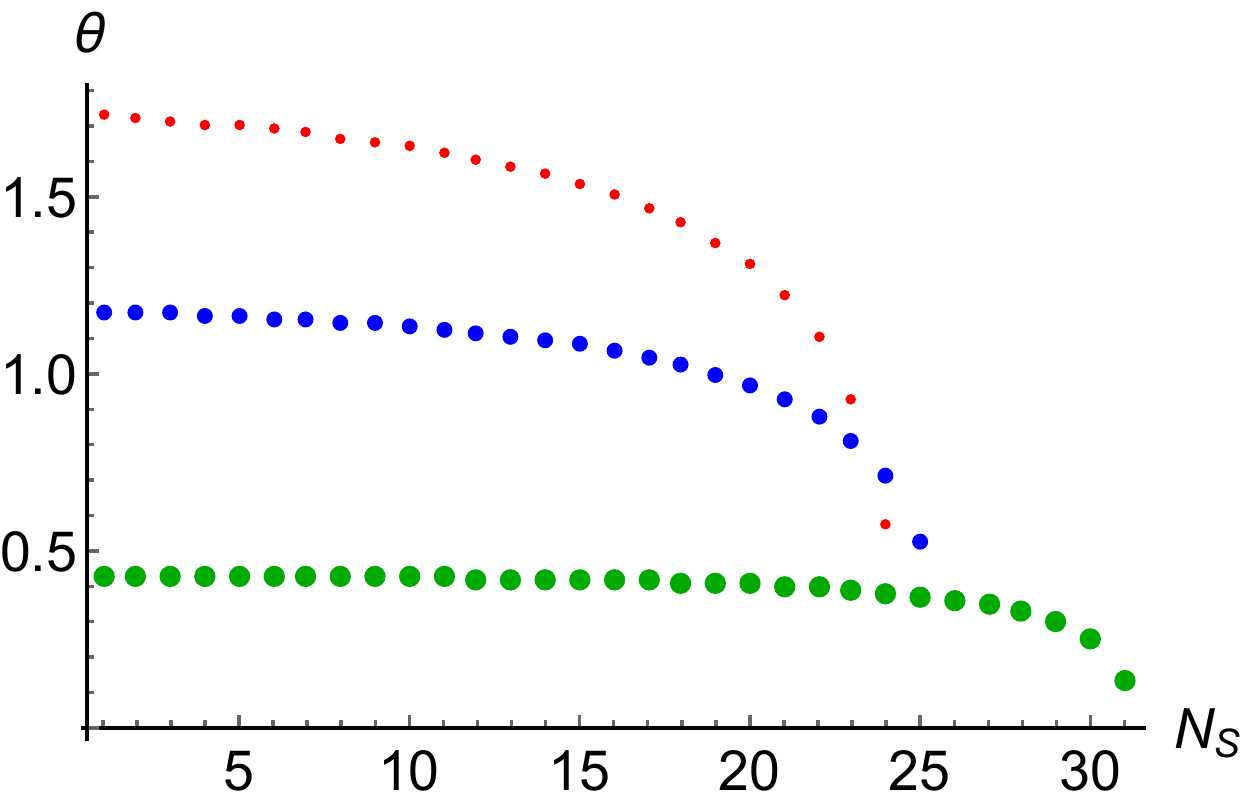}
\caption{\label{g3NsnoetasG3}We show the fixed point value for $g_3$ (upper panel) and the critical exponent $\theta$ (lower panel) as a function of $N_S$. The small red dots are for $G_3=G_4=1$, the medium blue ones for $G_3=G_4=3$ and the large green ones for $G_3=G_4=6$.}
\end{figure}

\begin{figure}[!here]
\includegraphics[width=0.45\linewidth]{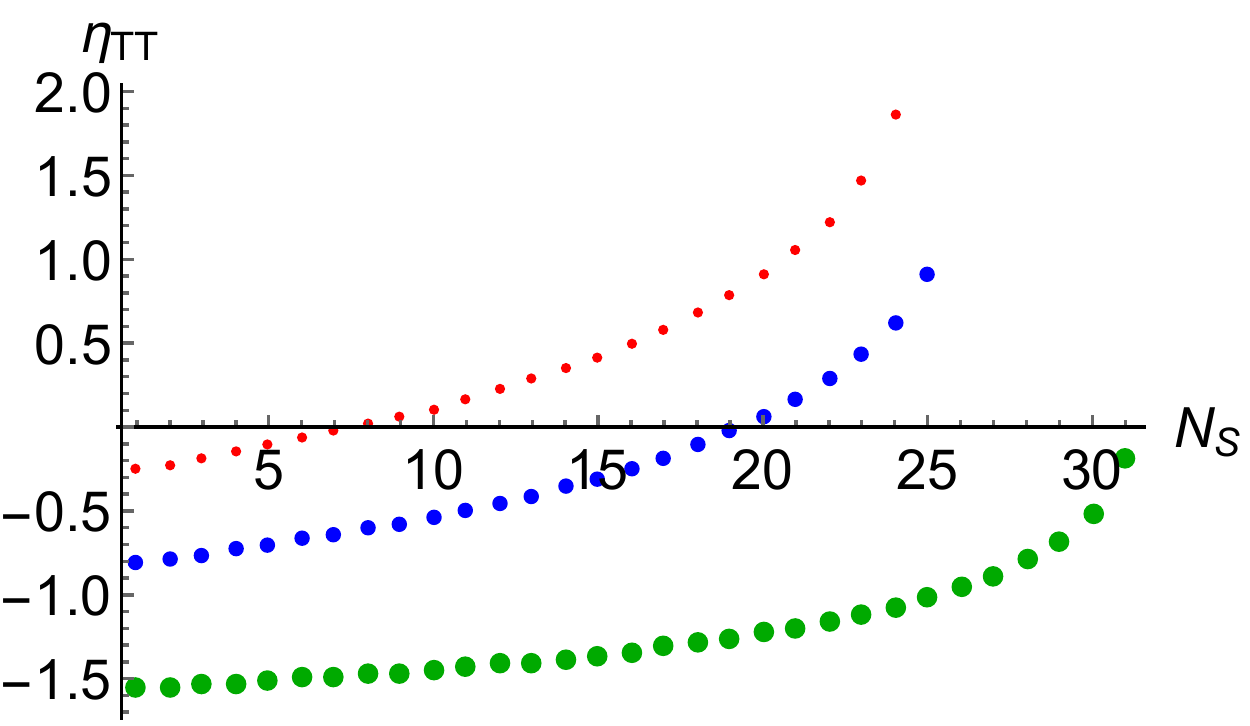}\quad
\includegraphics[width=0.45\linewidth]{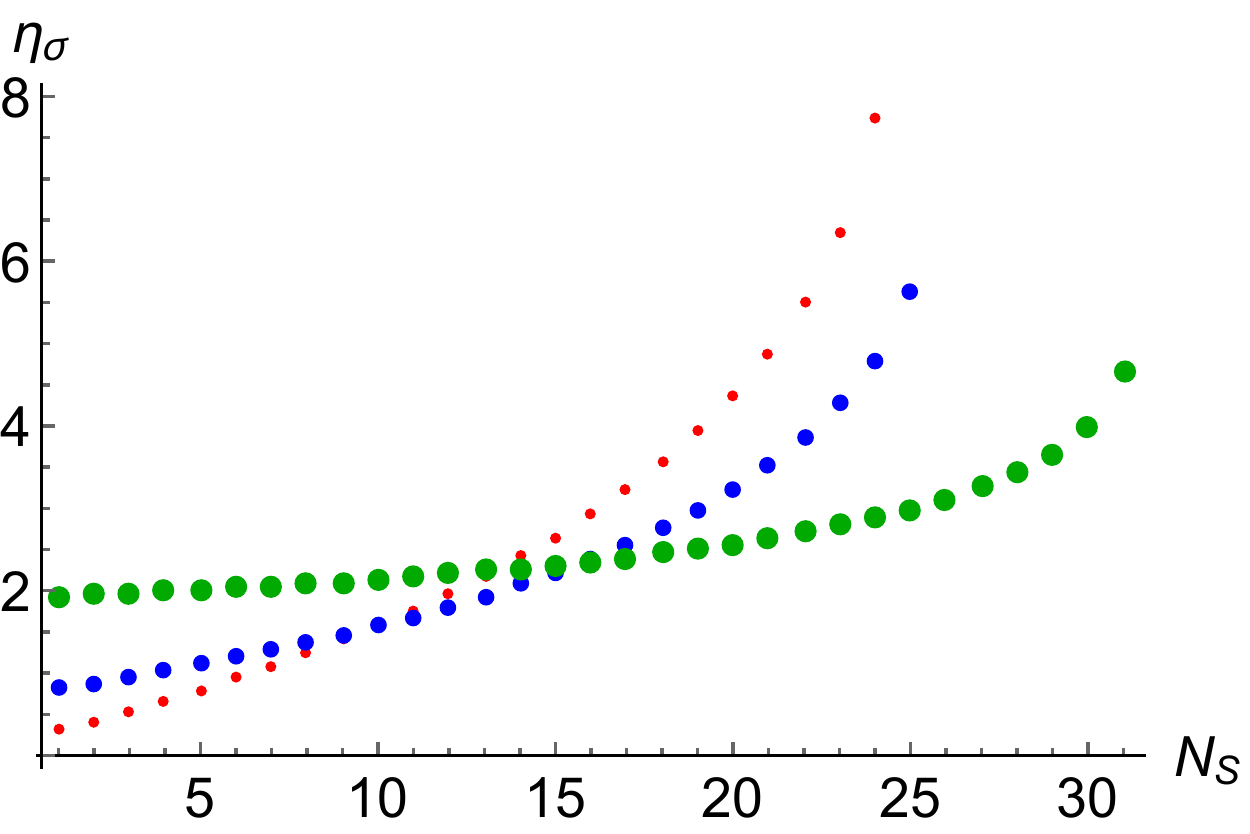}
\caption{\label{etaNsnoetasG3}We show the fixed point value for $\eta_{\rm TT}$ (left panel) and for $\eta_{\sigma}$ (right panel) as a function of $N_S$. The small red dots are for $G_3=G_4=1$, the medium blue ones for $G_3=G_4=3$ and the large green ones for $G_3=G_4=6$.}
\end{figure}

To investigate the stability of our results, we again distinguish the pure-gravity couplings from the gravity-matter couplings. We treat $G_4=G_3$ as an external parameter, and investigate the fixed point in $g_3$ as a function of $G_3$ and $N_S$. 
In particular, at smaller $G_3$, the value of $\eta_{\sigma}$ at small $N_S$ remains smaller. We observe that larger values of $G_3$ lead to a slower increase in the fixed-point value for $g_3$, and increase the value $N_S$ at which the fixed-point annihilation occurs, cf.~Fig.~\ref{g3NsnoetasG3}. While quantitative details change, the overall effect of increasing $N_S$ is similar to the previous approximation, cf.~Fig.~\ref{etaNsnoetasG3}. This suggests that our results will be qualitatively stable under extensions of the truncation including the running of the gravity-couplings as in \cite{Meibohm:2015twa}. On the other hand, a non-trivial interplay between the gravity-matter coupling and the pure-gravity coupling at large $N_S$ is not excluded, as the effect of varying $g_3$ on $G_{3\, \ast}$  remains to be studied.

\subsection{Fixed-point results including 
scalar anomalous dimension}\label{FPwithmatterincludingetas}

We now set all gravitational couplings and gravity-scalar couplings equal to the Newton coupling as defined from the gravity-matter interaction and include the anomalous dimension for the scalar, $\eta_S$.
This yields a new contribution to $\beta_{g_3}$, which now reads
\be
\beta_{g_3}=2 g_3 + 2 g_3 \eta_S- 
\frac{1}{6\pi}g_3^2\left(20-\frac{N_S}{4}+\frac{23}{60}\eta_S\right)
+\mathcal{O}(g_3^3).
\ee
Using
\be
\eta_S = \frac{7}{4 \pi}g_3 + \mathcal{O}(g_3^2)
\ee
equation (\ref{norma}) gets replaced by
\be
\beta_{g_3}= 2g_3+\frac{4+N_S}{24\pi}g_3^2
+\mathcal{O}(g_3^3).
\ee
Therefore, in this simplified form,
the terms of order $g_3^2$ in the beta function
are always positive, and by themselves would 
not produce a fixed point.
Fixed points appear when we consider also higher nonlinearities,
but their properties are not very stable.

For small numbers of scalars, both in the semi-perturbative 
approximation and using the full equations, we have,
in addition to the Gaussian fixed point,
also two non-trivial fixed points.
The first, which we call $FP_1$, has
negative $g_3$ and the second, which we call $FP_2$,
has positive $g_3$.
The fixed point $FP_1$ cannot be immediately
discarded on the basis of having a negative $g_3$.
While a negative value of the Newton coupling in the infrared is of course incompatible with observations, a negative fixed-point value is viable, as long as the RG flow can cross to $g_3>0$ towards the IR.
As the full beta function contains terms $\sim g_5$ etc., which imply $\beta_{g_3}\neq0$ at $g_3=0$, this situation is realized here. 

In the semi-perturbative approximation the solutions
of the fixed point equation can be written explicitly:
\be
g_{3*}=\frac{9\pi\left(60+15N_S\pm\sqrt{41904-2008N_S+45N_S^2}\right)}{1287-74N_S}
\ .
\ee
The solution $FP_2$ (which corresponds to the positive sign
in front of the root) is a growing function of $N_S$, with a simple pole
between $N_S=17$ and $18$ and asymptotes to $-135\pi/37$
for large $N_S$.
The solution $FP_1$ is a negative, smooth, monotonically increasing function of $N_S$ that asymptotes to zero.

The solutions of the full equations are more complicated to display
in closed form and are best studied numerically.
Let us start from the case $N_S=1$.
The properties of the fixed points in the two approximations
are listed in~tab.~\ref{FPtab3}.

\begin{table}[!here]
\begin{tabular}{cccccc}
approximation&$g_{3\,\ast}$& $\theta$& $\eta_{\rm TT}$&$\eta_{\sigma}$& $\eta_S$\\ \hline \hline
$FP_1$-s.p. & -8.67 & 3.42 & 2.34 &-1.92 &-4.83 \\ \hline
$FP_1$-full & -13.9 & 3.85 & 3.27 &-0.31 &-6.50 \\ \hline
$FP_2$-s.p. & 12.2 & 4.81 & -3.28 & 2.69 & 6.78 \\ \hline
$FP_2$-full & 53.4 & 11.7 & -5.21 &10.8 &25.2\\ \hline\hline
\end{tabular}
\caption{\label{FPtab3}Coordinates and critical exponents of the interacting fixed points at $N_S=1$, both in semi-perturbative and full approximations.}
\end{table}

The properties of the fixed points in the two approximations
are very roughly consistent, but $FP_2$ in the full approximation
has very large critical exponent and anomalous dimensions,
that put it far beyond the regime where our approximation is reliable.

The similarities end here, because the dependence on $N_S$
is quite different in the two approximations.
In the full calculation, $g_{3*}$ for $FP_2$,
as a function of $N_S$, is initially decreasing, 
has a minimum near $N_S=30$,
then increases again and the fixed point 
ceases to exist for $N_S>95$.
It switches from UV attractive to repulsive near $N_S=47$
and there is no value of $N_S$ for which the critical exponents
have reasonably small values.
This is very different from its behavior in the semi-perturbative
approximation.

Also $FP_1$ in the full calculation behaves very differently from
the semi-perturbative approximation:
instead of increasing steadily to zero, it decreases monotonically
and diverges to $-\infty$ for $N_S$ just above $46$.
Some of the anomalous dimensions have more reasonable values
but there is no $N_S$ for which they are all small.
Thus $FP_1$ has slightly better chances of being
a true fixed point, but in the present approximations
it cannot be reliably assessed.

\begin{figure}[!here]
\includegraphics[width=0.45\linewidth]{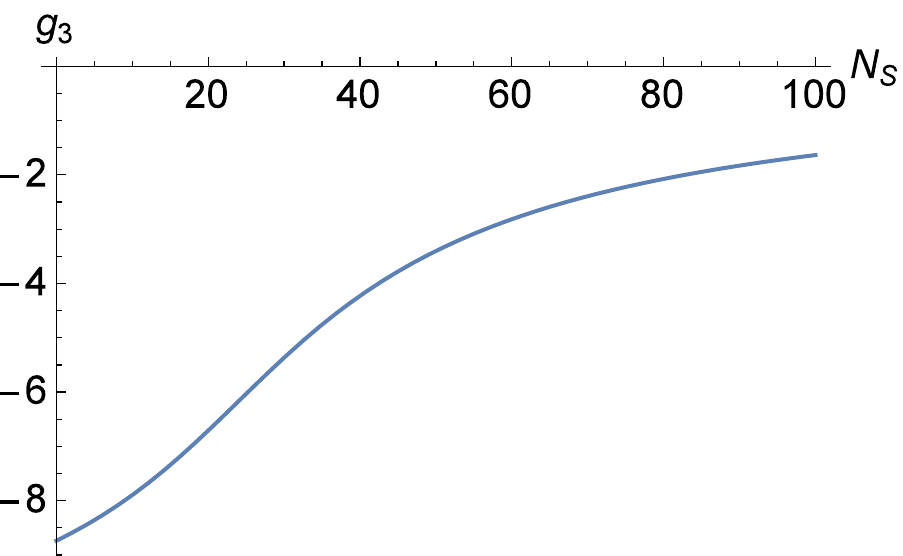} \quad \includegraphics[width=0.45\linewidth]{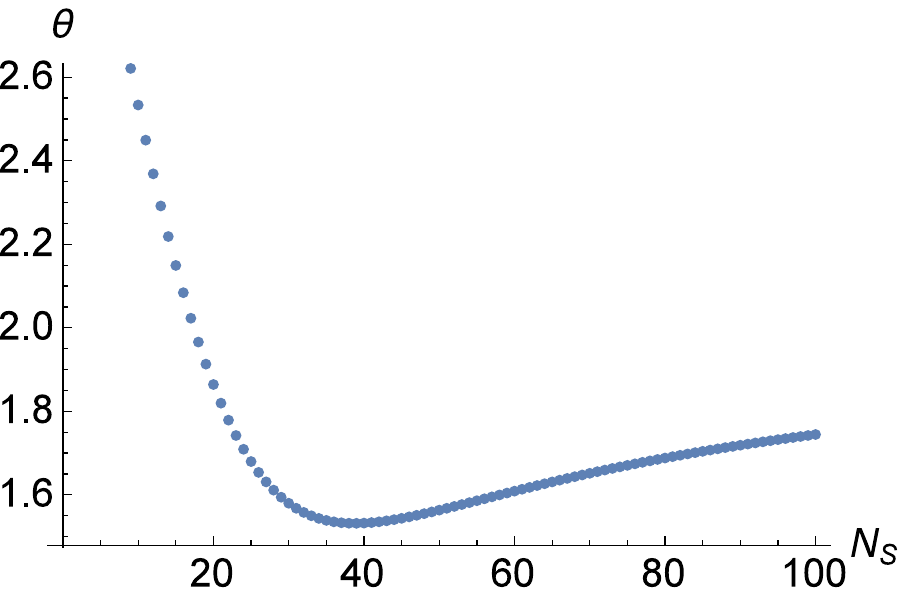}\newline
\includegraphics[width=0.45\linewidth]{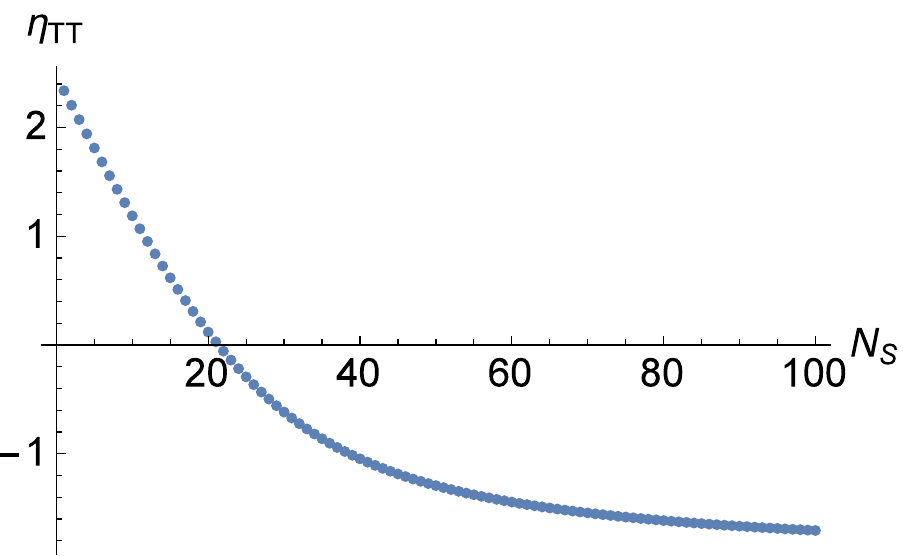} \quad \includegraphics[width=0.45\linewidth]{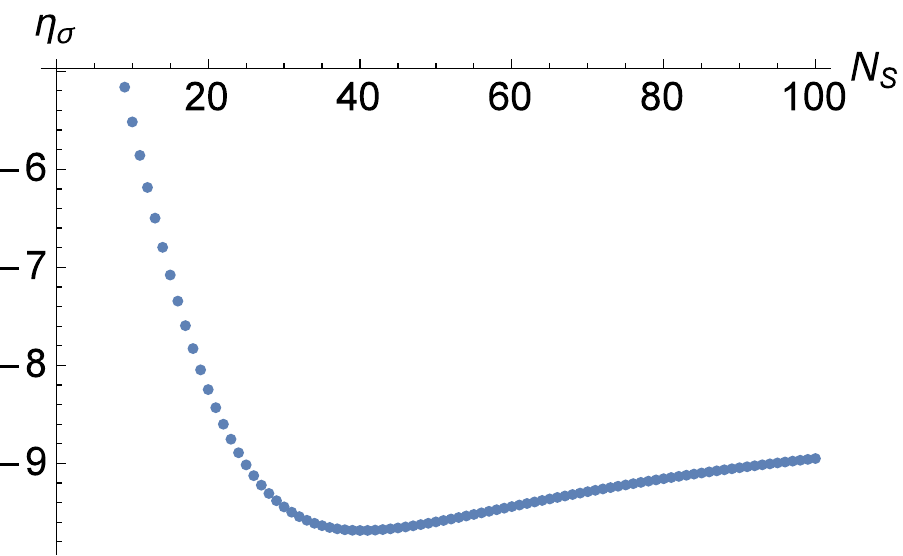}\newline
\includegraphics[width=0.45\linewidth]{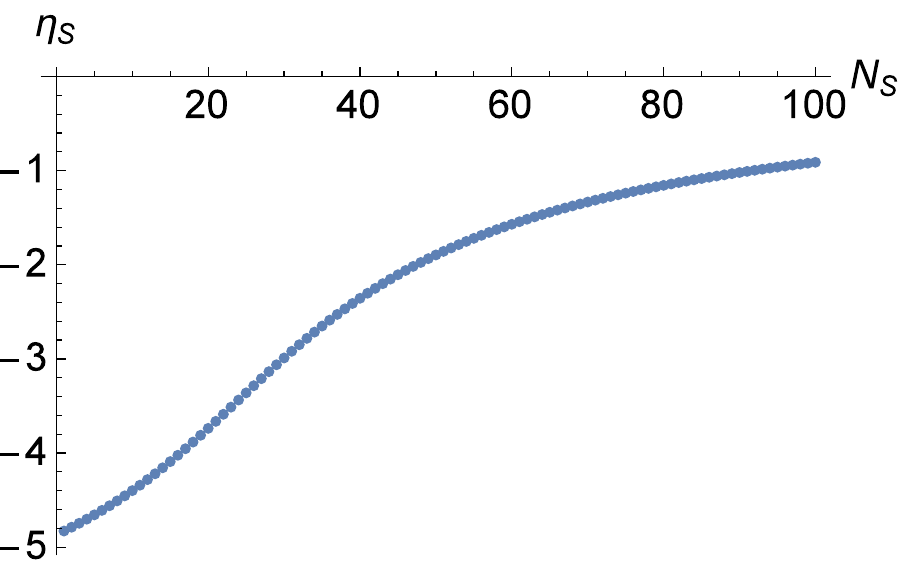} 
\caption{\label{FPNsequalcouplings} We show results in the semi-perturbative approximation: We plot the fixed-point value for $g_3$ as a function of $N_S$ (left upper panel) and the value of the critical exponent $\theta$ (right upper panel), as well as the anomalous dimensions.}
\end{figure}

We observe that already in the case $N_S=1$ 
the anomalous dimensions of the graviton modes 
have opposite signs compared to the case when we neglected $\eta_S$,
and become rather large, as a consequence of a rather large absolute value of $g_3$. 
The critical exponent also differs considerably from the pure-gravity case. 
As a consequence, when we take into account the effect of
$\eta_S$, it becomes hard to identify either of the fixed points
with the one that we found for $N_S=0$.

These results could lead to different conclusions:
the approximation $G=g$ might not be particularly reliable 
beyond small $N_S$, or, more likely, our current estimate 
for the scalar anomalous dimension needs improvement, 
and the results for $\eta_S=0$ might be closer to the correct result.

The significant change in the fixed-point properties from the pure-gravity case arises from the scalar anomalous dimension. 
Accordingly, the perturbative approximation, where $\eta_{TT}=\eta_{\sigma}=\eta_S=0$ 
features a fixed point at $g_3=2.57$ with $\theta=2$ 
for all values of $N_S$. 
As $\eta_S$ has such a significant effect on the existence and properties of fixed points, it is important to understand 
whether our truncation already captures all major operators 
that determine $\eta_S$. Recall that metric fluctuations 
induce non-vanishing momentum-dependent 
matter self-interactions, e.g., of the form $\left(g^{\mu \nu} \partial_{\mu} \phi \partial_{\nu}\phi\right)^2$ \cite{Eichhorn:2012va}. As soon as these couplings 
are non-zero, they yield a nonvanishing contribution to $\eta_S$. 
Our current truncation does not include these effects. 
We therefore conjecture that the results could improve 
once these further operators are included. 
By a simple count of modes, $N_S=1$ should not 
dramatically change the results, and the case $N_S=1$ 
should still feature a fixed point 
with properties similar to the pure-gravity one, 
just as exhibited by the approximation $\eta_S=0$.
We tentatively suggest that the calculation with $\eta_S=0$,
which shows a destablizing effect setting in at $N_S\gg1$, 
might capture the full dynamics more accurately than our current estimate with $\eta_S \neq 0$.

\subsection{Background beta functions}
The background couplings can only appear on the right-hand side of beta functions through a trivial scaling, for instance, $\beta_{\bar{G}} \sim \bar{G}^2$, as the prefactor of the curvature term in the Einstein-Hilbert action is $\frac{1}{16 \pi \bar{G}}$. All couplings that appear on the right-hand side from either propagators or vertices are always fluctuation field couplings. Accordingly, $\eta_{\rm TT}$, $\eta_{\sigma}$ and $\eta_S$, which will appear on the right-hand side of $\beta_{\bar{G}}$, depend on the fluctuation-field couplings $g_3$, $g_4$, $g_5$, $G_3$, $G_4$ only. 
Thus we obtain the following matter contribution to the $\beta$ function for the background Newton coupling:
\be
\beta_{\bar{G}}\Big|_{\rm scalar}=\frac{\bar{G}^2}{24 \pi}N_S (4-\eta_S).
\ee

Following \cite{Percacci:2015wwa},
\bea
\beta_{\bar{G}}=2 \bar{G} 
- \frac{\bar{G}^2}{\pi} \left(\frac{15}{8}
- \frac{5 \eta_{\rm TT}}{18 }
+ \frac{\eta_{\sigma}}{24} 
-  \frac{N_S}{24} (4 - \eta_{S})\right).
\eea

If we set $\eta_{\rm TT}= \eta_{\sigma}=0=\eta_S$, we obtain $N_S=45$ as the maximal number of scalars before the fixed point in $\bar{G}$ diverges.

If we now use the approximation of equating all fluctuation couplings in the expressions for the anomalous dimensions, as in Sec.~\ref{FPwithmatterincludingetas}, we obtain a fixed point at $\bar{G}_{\ast}=12.14$, with critical exponent $\theta=2$, for $N_S=1$  at the fixed point $g_{\ast}=-13.9$. 
 The location of a possible bound depends on the assumptions for the couplings $g_4, g_5$, $G_3, G_4$. For the case $g_4=0=g_5$, $G_3=0=G_4$, the continuation of the pure-gravity fixed point in the background coupling ceases to exist beyond $N_S=9$.
 
For the approximation where we set $\eta_S=0$, we obtain a bound at $N_S=12$, which is close to the bound from the fluctuation coupling, at $N_S=14$.

Note that fixed-point values for the background couplings are affected by a strong regulator dependence: As the background metric enters the regulator function, there are contributions to all beta functions of background couplings that are due to the regularization only, and are unphysical in that sense, see also \cite{Folkerts:2011jz,Litim:2002ce, Litim:2002hj,Bridle:2013sra}.
Thus, even divergences in background couplings might turn out to be compatible with a model that is asymptotically safe in a physical sense, i.e., where all physical quantities have a well-behaved UV limit. To understand on which couplings a fixed-point requirement must be imposed, and which couplings may even diverge, one must investigate physical observables. As this is clearly beyond the scope of the present work, we conclude that the most conservative assumption is that \emph{all} couplings must feature fixed points, and therefore divergences in the background couplings are not acceptable, even if the fluctuation couplings are well-behaved. Interpreted along these lines, the one-loop approximation, where we set $\eta_{\rm TT}=\eta_{\sigma}=0=\eta_S$ everywhere, and only the background Newton coupling depends on $N_S$, would suggest that there could be an upper limit of scalars that is compatible with a viable fixed point.\newline\\

\section{Conclusions}
In this work, we discuss how setting up a Renormalization Group flow for gravity-matter systems, in order to investigate the viability of the asymptotic safety scenario for gravity and the Standard Model, necessitates a distinction between couplings of matter to the background metric and couplings to fluctuations of the metric. We take a first step in disentangling the scale-dependence of the different couplings by studying the running of a vertex at which two scalar fields interact with the metric fluctuation field. From this vertex we define an avatar of the Newton coupling, $g_3$.
We observe that the so-defined Newton coupling features an interacting fixed point at $N_S=0$ where only metric fluctuations drive the Renormalization Group flow. The universal critical exponent at this fixed point is close to that of other approximations and definitions. We consider this rather strong evidence for the asymptotic safety scenario  in the pure-gravity case, that complements previous results. We emphasize that this is the first evidence for asymptotic-safety in gravity-matter interactions, as all previous results related to the Newton coupling defined from background or fluctuation gravitational interactions. Our result is therefore a new hint that formulating an asymptotically safe theory of gravity and matter could be phenomenologically viable.\newline

We also investigate the anomalous dimensions for different components of the graviton, and find a large negative anomalous dimension $\eta_{\rm TT}$ for the TT mode, which is not too far from the single-metric approximation $\eta_N=-2$.  Such a large negative anomalous dimension implies a strongly UV suppressed propagator of the form 
$p^{-2+\eta_{TT}}$. The corresponding propagator in real space is reminiscent of a lower-dimensional setting. Thus our result is in line with other indications for some form of dynamical dimensional reduction in asymptotically safe gravity   \cite{Lauscher:2001ya, Lauscher:2005qz,Reuter:2011ah,Rechenberger:2012pm,Calcagni:2013vsa}, however see also \cite{D'Odorico:2015lhd}.

On the other hand, the  $\sigma$-anomalous dimension has the opposite sign. This suggests that different tensor structures in gravity exhibit different running - reminiscent of the difference between the transverse and longitudinal gluons in the infrared regime in Yang-Mills theory in Landau gauge. 
This result is an indication that one should also disentangle the flow of different tensor structures at the level of the vertices.
We moreover observe a difference to results in the linear parametrization, where the anomalous dimensions are typically smaller in  absolute value.

 Within functional Renormalization Group flows, it is never possible to find a finite-dimensional closed truncation, as higher-order couplings always couple back into the flow of lower-order ones. For instance, the flow of the coupling $g_3$ depends on the higher-order coupling $g_5$ through a tadpole diagram. Thus, one should evaluate the flow for $g_5$, which itself depends on $g_7$ and so on, necessitating some approximation to close the truncation. Typical choices in the literature include setting higher-order couplings to zero, or equating them to lower order ones. In our results, we explicitly keep the dependence on all couplings that enter $\beta_{g_3}$, enabling us to study the reliability of different approximations:
We observe that the choice of approximation for those couplings for which no beta function is determined explicitly, quantitatively alters the properties of the fixed point: By treating, e.g., the pure-gravity couplings $G_3$ and $G_4$ as external parameters, we observe that the fixed point in $g_3$ persists for all values of these couplings that we have investigated, but, e.g., anomalous dimensions and the critical exponent change. It is reassuring that the existence of a fixed point does not  depend on making specific choices for couplings for which no beta function is determined, as of course all results in the literature make specific choices for these couplings. On the other hand, our investigation of the dependence of fixed-point properties on these choices highlights that quantitatively more precise results require
 more elaborate truncation schemes which disentangle some of these couplings.

In the case $N_S>0$, scalar fluctuations have a significant impact. If we work in the approximation where we set the anomalous dimension for the scalar matter field to zero, $\eta_S=0$, we observe that matter fluctuations  have a destabilizing effect on the gravitational fixed point, and move it toward larger values.  This observation is in accordance with the general scenario discussed at the level of the background couplings in \cite{Dona:2013qba, Dona:2014pla}, where it is argued that the inclusion of dynamical matter degrees of freedom will impact the microscopic dynamics for gravity, and an increasing number of scalars leads to a growth of the fixed point value for the Newton coupling. In \cite{Meibohm:2015twa} a similar behavior is found for the fluctuation coupling $G_3$. 
The increasing fixed-point value leads to an increase in the anomalous dimension for the graviton. 
As our regularization scheme requires $\eta<2$ for all anomalous dimensions, the region of $N_S \geq 14$ requires a re-investigation with a different regularization scheme and/or significantly larger truncations. 
We should therefore take care when interpreting our present results. Keeping in mind this word of caution, we observe that scalar matter seems to have a significant effect on an interacting fixed point and could potentially destabilize it.  On the other hand, it is reassuring to observe that $N_S=4$, which is of course the phenomenologically most relevant case as it corresponds to the number of scalar fields in the Standard Model, admits a gravitational fixed point in our setting, again in line with results in \cite{Dona:2013qba, Dona:2014pla, Meibohm:2015twa}.

Including a scalar anomalous dimension $\eta_S$ leads to a very significant change of the fixed-point properties already at $N_S=1$. 
Based on a mode-counting argument, we 
expect that this strong effect of a single scalar field only arises within our truncation, and such significant effects of matter should not be expected for small $N_S$. We also identify a direction in which an extension of our truncation would potentially lead to $\mathcal{O}(1)$ changes of the anomalous dimension for scalars: Quantum-gravity effects induce non-vanishing momentum-dependent self-interactions for scalars \cite{Eichhorn:2012va}, which will couple into $\eta_S$ via a tadpole diagram.
If we take into account quantum-gravity induced matter self-interactions, not only $\eta_S$ will change. Among the diagrams contributing to $\beta_{g_3}$, there will also be one that features a closed scalar loop, and thus yields a contribution that scales with $N_S$. Within our present approximation, there is no such explicit contribution, and the $N_S$ dependence only arises through the anomalous dimensions.

 Moreover, it will be interesting to understand the parametrization and gauge-dependence of our result. As the matter-gravity vertices differ signficiantly between, e.g., Landau gauge and linear parametrization on the one hand, and unimodular gauge and exponential parametrization on the other hand, the structure of the beta function for $g_3$ as well as $G_3$ could be significantly different. Thus, it will be interesting to understand whether the observations in the present work as well as in \cite{Meibohm:2015twa} persist, if the system $\beta_{g_3}, \beta_{G_3}$ is analyzed in the two different parametrizations. In particular, it will be interesting to see whether the sign of the contribution $\sim N_S$ to $\beta_{g_3}$ and $\beta_{G_3}$ depends on the choice of parametrization.

\emph{Acknowledgements}
We thank Manuel Reichert and particularly Jan M. Pawlowski for insightful discussions. The work of A.E. is supported by an Imperial College Junior Research Fellowship.

\end{document}